\documentclass{aa}

\usepackage{graphicx}
\usepackage{txfonts}
\usepackage{bm}
\usepackage{hyperref}
\usepackage{comment}

\begin{document}

   \title{Detecting galaxy -- 21-cm cross-correlation during reionization}

   \author{Samuel Gagnon-Hartman\thanks{email:samuel.gagnonhartman@sns.it}\inst{1}
          \and
          James E. Davies\inst{1}
          \and
          Andrei Mesinger\inst{1,2}
          }

   \institute{Scuola Normale Superiore di Pisa,
              Piazza dei Cavallieri 7, 56126 Pisa, Italy
             \and 
            Centro Nazionale ``High Performance Computing, Big Data and Quantum Computing'', Via Magnanelli, 
            40033 Casalecchio di Reno, Italy\\
            }

   \date{Received 10 March 2025 / accepted 13 May 2025}
 
\abstract{
The cosmic 21-cm signal promises to revolutionize studies of the Epoch of Reionization (EoR).  Radio interferometers are aiming for a preliminary, low signal-to-noise (S/N) detection of the 21-cm power spectrum.  
Cross-correlating 21-cm with galaxies will be especially useful in these efforts, providing both a sanity check for initial 21-cm detection claims and potentially increasing the S/N due to uncorrelated residual systematics. 
Here we self-consistently simulate large-scale (1 Gpc$^3$) galaxy and 21-cm fields, computing their cross-power spectra for various choices of instruments and survey properties.  We use $1080$h observations with SKA-low AA* and HERA-350 as our benchmark 21-cm observations.  We create mock Lyman-$\alpha$ narrow-band, slitless and slit spectroscopic surveys, using benchmarks from instruments such as Subaru HyperSupremeCam, Roman grism, VLT MOONS, ELT MOSAIC, and JWST NIRCam.  We forecast the resulting S/N of the galaxy--21-cm cross-power spectrum, varying for each pair of instruments the galaxy survey area, depth, and the 21-cm foreground contaminated region of Fourier space.
We find that the highest S/N is achievable through slitless, wide-area spectroscopic surveys, with the proposed Roman HLS survey resulting in a $\sim55\sigma$ ($\sim13\sigma$) detection of the cross-power with 21-cm as observed with SKA-low AA* (HERA-350), for our fiducial model and assuming $\sim$500 sq. deg. of overlap.
Narrow-band dropout surveys are unlikely to result in a detectable cross-power, due to their poor redshift localization. Slit spectroscopy can provide a high S/N detection of the cross-power for SKA-low AA* observations.  Specifically, the planned MOONRISE survey with MOONS on the VLT can result in a $\sim3\sigma$ detection, while a survey of comparable observing time using MOSAIC on the ELT can result in a $\sim4\sigma$ detection. Our results can be used to guide survey strategies, facilitating the detection of the galaxy--21-cm cross-power spectrum.
}


\maketitle

\section{Introduction}

% P1
The 21-cm signal from neutral hydrogen is arguably 
the most promising probe of the Cosmic Dawn (CD) 
and the Epoch of Reionization (EoR): two of the last 
remaining frontiers in modern cosmology. 
The signal holds imprints of the physics of the first galaxies 
(e.g., \citealt{mirocha19,park2019}), the nature of dark 
matter (e.g., \citealt{clark18, vd21, facchinetti24}), 
and perhaps even the role of dark energy (e.g., 
\citealt{adi24}) during this crucial 
period in the development of the Universe.

% P2
A number of challenges riddle the path toward a measurement of 
the 21-cm power spectrum. The interferometer arrays used 
in tomographic experiments introduce numerous systematic effects into their 
measurements, all of which must be carefully characterized and accounted 
for (e.g., \citealt{kern20,rath24}). Furthermore, the highly redshifted 21-cm signal
occupies the same radio band as Galactic synchrotron radiation, which 
overwhelms the cosmic signal by several orders of magnitude. There is also terrestrial contamination to contend with, including radio-frequency interference and the Earth's ionosphere (e.g., \citealt{liushaw20}).

Owing to these challenges, any claim of a detection of the 21-cm power spectrum
will require some form of confirmation in order to gain the trust 
of the scientific community. The most effective method to bolster 
confidence in a cosmic 21-cm detection is to correlate it with a signal of known cosmic origin.

In addition to building confidence, a cross-correlation signal could 
even increase the signal-to-noise (S/N) of a detection, because the foregrounds 
of the two signals are probably not correlated. Several low-redshift studies have indeed confirmed measurements of the 21-cm signal through cross-correlation with galaxy/quasi-stellar object surveys spanning the same volume (e.g., the Green Bank 
Telescope \citealt{masui13}, Parkes Radio Telescope \citealt{anderson18}, MeerKAT \citealt{cunnington23}, CHIME  
\citealt{amiri23b,amiri23}). Depending on its nature, the cross-power spectrum could also probe the interaction between 
galaxies and the intergalactic medium (IGM), yielding 
complementary information to the auto-power spectra (e.g., 
\citealt{hutter23, moriwaki24}). 

\begin{figure*}
    \resizebox{\hsize}{!}
    {\includegraphics[width=\textwidth, trim=10 70 10 100, clip]{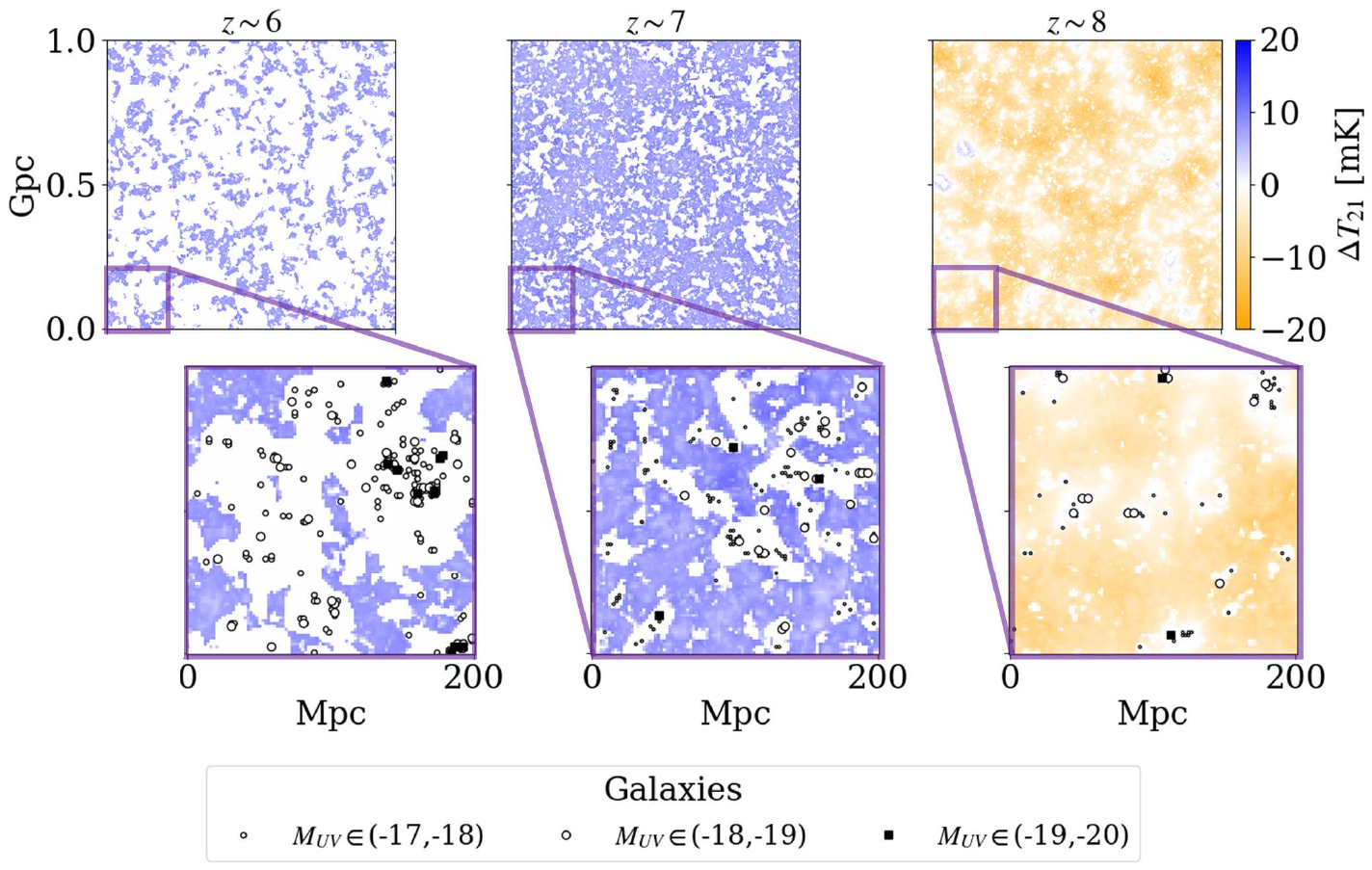}}
    \caption{An example 21-cm brightness temperature field at shown
    at redshifts $6$, $7$, and $8$ (left to right) 
    with galaxies overlaid on zoomed insets (black). 
    Detectable galaxies at these redshifts preferentially reside in ionized regions, 
    with the brightest, most biased galaxies residing in the largest ionized bubbles. The top row shows slices of the full gigaparsec volume used in this study, while the bottom row zooms into a $200$ Mpc corner for clarity. Both the brightness temperature and galaxy fields were generated 
    simultaneously using \texttt{21cmFASTv4}.}
    \label{fig:cutout}
 \end{figure*}

For the  EoR / CD 21-cm signal at $z \gtrsim 5$, 
there are three main candidates for cross-correlation: 
cosmic backgrounds, intensity maps, and galaxy surveys, 
each with its own benefits and challenges.
The cosmic microwave background (CMB) is a very well 
established signal and can cross correlate with the 21-cm 
signal through secondary anisotropies such as the 
kinetic Sunyaev-Zel'dovich effect (e.g., 
\citealt{reichardt21,adachi22}). It is an 
integral (2D) signal, however, and the lack of line-of-sight modes 
causes them to correlate only weakly with the cosmic 
21-cm signal (e.g., \citealt{lidz09,mao14,ma18,vrbanec20,laplante22,sun25}). 
On the other hand, line-intensity maps (LIMs) of 
other emission lines are expected to correlate very strongly 
with the cosmic 21-cm signal. Most LIMs target 
metals in the interstellar medium, however, such as CO and CII, 
whose corresponding luminosities and achievable S/N of the cross-correlation with 21-cm are very uncertain at $z\gtrsim5$ (e.g., \citealt{crites14,cooray16,kovetz17,parshley18,heneka21,moriwaki21,yang22,pallottini23,fronenberg24}).

While LIM experiments will take time to mature, the galaxy mapping toolkit is not only well established 
but booming at the time of writing. A plethora of 
ground and space instruments (e.g., the James Webb 
Space Telescope, the Subaru Telescope, the 
Atacama Large Millimeter Array) are currently detecting galaxies 
deep in the EoR. While time-consuming 
to produce, spectroscopic maps probe the line-of-sight 
modes crucial for cross-correlation with the 
cosmic 21-cm signal while avoiding the challenges associated with LIM 
\citep{sobacchi16,hutter17,kubota20,vrbanec20,la_plante23}. 
As the drivers of reionization, galaxies are tightly correlated 
with ionized bubbles (see Figure \ref{fig:cutout}).  
Given a sufficient number of galaxies, we should 
thus detect a strong (anti-) correlation with the 
21-cm emitting neutral IGM during the EoR.

We quantify the detectability of the cross-power 
spectrum between the 21-cm signal and a galaxy survey 
spanning the same volume. To do this, we self-consistently simulate 
large-scale ($1$ Gpc) maps of the 21-cm brightness temperature 
and the underlying galaxy field. Using empirical relations that 
include scatter, we made mock observations of galaxy maps with 
selection based on Lyman-$\alpha$ and other nebular lines. 
We quantified the S/N of the galaxy--21-cm cross-power 
as a function of galaxy survey depth, angular extent,
and spectral resolution, as well as the level 
of foreground contamination of the cosmic 21-cm signal. 
Based on these estimates, we forecast the S/N for 
current and upcoming telescopes, including HERA and 
SKA for 21-cm, and Subaru, Roman, MOONS and MOSAIC for 
galaxy maps. We conclude with specific recommendations 
for surveys seeking to maximize the probability of 
detecting the galaxy--21-cm cross-power spectrum.

The paper is organized as follows. 
In Section \ref{sec:sim} we present our 
cosmological simulations. In Section \ref{sec:det} 
we convert our simulation outputs into a mock signal. 
This involves various assumptions about 
instrumental effects and sources of systematic uncertainty.
In Section \ref{sec:exp}  we discuss fiducial instruments 
and the survey choices we varied to quantify the detectability of 
the cross-power spectrum.  We present our results 
in Section \ref{sec:disc}. We conclude in Section 
\ref{sec:conc} with specific recommendations for surveys 
seeking to detect the cross-power. We assume a 
standard $\Lambda$CDM cosmology with parameters consistent 
with the latest {\it Planck} estimates throughout, 
wherein $H_0=67.66$ km s$^{-1}$ Mpc$^{-1}$, 
$\Omega_{m}=0.30966$, $\Omega_{b}=0.04897$, 
$\Omega_k=0.0$, $\Omega_\Lambda=0.68846$, and $\sigma_8=0.8102$. 
Unless stated otherwise, all lengths are quoted in comoving units. 

\section{Simulating the cosmic signal}
\label{sec:sim}

The cosmic signal in this study refers to two fields: 
the 21-cm brightness temperature and an overlapping galaxy map. 
To simulate the two fields, we used the latest version of the 
public code \texttt{21cmFAST} \citep{mesinger07, mesinger2011}. 
We then postprocessed the cosmic signals, added telescope 
noise, removed foreground-contaminated regions, and implemented 
selection criteria for galaxy surveys. This section is split 
into two subsections; the first subsection summarizes how the latest version of \texttt{21cmFAST} computes cosmological fields and the 
second subsection discusses our procedure for applying the appropriate observational selection criterion to our simulated galaxy maps.

\subsection{Cosmological simulations}

We used the latest public release, \texttt{21cmFASTv4}\footnote{\url{https://github.com/21cmFAST/21cmFAST/tree/v4-prep}} 
\citep{davies25}, which allows for greater flexibility in characterizing high-redshift galaxies. 
We describe the simulation below with a focus on the new galaxy source model. 
For further information, we refer to \cite{davies25}.

\texttt{21cmFAST} creates a 3D realization of the matter field at $z=30$, 
evolving it to lower redshifts through second-order Lagrangian perturbation theory (2LPT). We used simulation boxes 
that are $1$ Gpc on a side, 
with initial conditions sampled on a $1100^3$ grid. 
This was subsequently down-sampled to a cell size of $2$ Mpc. 
In the previous default version of \texttt{21cmFAST}, 
the source fields driving heating and ionization 
were calculated on filtered versions of the evolved 
density grid according to the excursion set algorithm 
\citep{mesinger2011}. In the default settings, galaxies were not 
localized on the Eulerian grid. 
While \texttt{21cmFAST} has the option to use a discrete halo 
field via a Lagrangian halo finder named \texttt{DexM}, 
this algorithm samples over scales with a minimum mass that is 
set using the size of the grid cell. This renders it prohibitively 
slow for halos with a mass of $M_h\leq10^{10}$ M$_\odot$ \citep{mesinger07}.

The new model of \citet{davies25} uses a combination of 
Lagrangian halo-finding and coarse time-step merger trees 
to rapidly build 3D realizations of dark matter halos throughout 
the EoR/CD. These halos are then populated by galaxies whose 
properties are sampled from empirical relations conditional on their halo mass and other properties. \texttt{21cmFASTv4} therefore 
explicitly accounts for stochastic galaxy formation and 
generates 3D lightcones of galaxy properties (e.g., stellar 
mass and star formation rate) alongside the corresponding IGM 
lightcones. Furthermore, these halos are tracked self-consistently 
across cosmic time. 

We assigned a stellar mass and star formation rate 
to each galaxy by sampling from a log-normal distribution 
conditioned on the mass of the host halo:
\begin{equation}
    \log_{10}M_*\sim\mathcal{N}(\log_{10}\mu_*(M_h), \sigma_*),
\end{equation}
\noindent where $\sigma_*$ is a redshift-independent free parameter
of the model, and the mean so-called stellar-to-halo mass relation (SHMR), is given by
\begin{equation}
    \mu_* = f_{*,10}\left(\frac{(M_{p1}/M_{p2})^{\alpha_*}}{(M_h/M_{p2})^{-\alpha_*}+(M_h/M_{p2})^{-\alpha_{*2}}}\right) M_h\exp\left(-\frac{M_{\textrm{turn}}}{M_h}\right)\frac{\Omega_b}{\Omega_m}.
\end{equation}
\noindent Here, $\Omega_b/\Omega_m$ is the mean
baryon fraction, and the fraction of galactic gas in stars 
was modeled as a double power law
defined by pivot masses ($M_{p1}$, $M_{p2}$) and slopes 
($\alpha_{*}$, $\alpha_{*2}$) (e.g., \citealt{mirocha2017}). 
Most 21-cm simulations neglect the high-mass turnover in the 
stellar-to-halo mass relation as these galaxies contribute very little 
to the total emissivity (e.g., \citealt{bouwens15re,park2019,gillet20,qin24}). 
However, these rare massive galaxies are bright enough to 
be observed at high redshifts are therefore fundamental to our study.

The star formation rate (SFR) of a galaxy was similarly sampled 
from a log-normal distribution conditioned on the stellar mass of the galaxy,
\begin{equation}
    \log_{10}\dot{M}_*\sim\mathcal{N}(\log_{10}\mu_{\textrm{SFR}}(M_h), \sigma_{\textrm{SFR}}),
\end{equation}
\noindent with the mean (typically referred to as the star-forming main sequence; SFMS)
\begin{equation}
    \mu_{\textrm{SFR}} = \frac{M_*}{t_* H^{-1}(z)}.
\end{equation}
\noindent Here, $t_*$ is a free parameter denoting a characteristic star formation timescale in units of the Hubble time $H(z)^{-1}$. Following \cite{davies25}, we decreased the dispersion around the SFMS with increasing stellar mass
\begin{equation}
\sigma_\textrm{SFR}=\textrm{max}\left[\sigma_{\textrm{SFR,lim}},
    \sigma_{\textrm{SFR,idx}}\left(\frac{M_*}{10^{10}\textrm{ M}_\odot}+\sigma_{\textrm{SFR,lim}}\right)\right].
\end{equation}
\noindent These fiducial choices were motivated by results from hydrodynamic simulations (e.g., FirstLight \citealt{ceverino18}; 
ASTRID \citealt{bird22,davies23}; SERRA \citealt{pallottini23}).

The nonionizing UV luminosity primarily from massive young stars was 
taken to be directly proportional to the star formation rate,
\begin{equation}
    \dot{M}_*(M_h,z)=
    \kappa_{\text{UV}}\times L_{\text{UV}},
\end{equation}
\noindent where we assumed the conversion factor 
$\kappa_\text{UV}=1.15\times10^{-28}
M_\odot \text{ yr}^{-1}/\text{erg s}^{-1}
\text{Hz}^{-1}$ (e.g., \citealt{sun16}). 
Throughout this work, we refer to UV luminosities in units of AB 
magnitudes, which we computed via the standard relation (e.g., 
\citealt{oke83})
\begin{equation}
    \log_{10}\left(\frac{L_{\text{UV}}}
    {\text{erg s}^{-1}\text{Hz}^{-1}}\right)
    =0.4\times (51.63-M_{\text{UV}})\label{eq:mab} ~ .
\end{equation}
The fraction of ionizing radiation that escapes the galaxy is 
characterized by the escape fraction, $f_\textrm{esc}$, 
whose value is notoriously elusive. Hydrodynamic 
simulations (e.g., \citealt{kimm15,xu17,barrow17,kostyuk23}), 
direct observations of low-redshift galaxies (e.g., \citealt{izotov16,grazian17,steidel18,pahl23}), 
as well as inference from EoR observations (e.g., \citealt{nikolic23,qin24,chakraborty24}) 
result in very different estimates of $f_\textrm{esc}$ and its dependence on galaxy properties. To avoid introducing unmotivated 
complexity into our model, we adopted the best-fit value of the 
population-averaged escape fraction, $\bar{f}_{\textrm{esc}}=0.05$, 
from \citet{nikolic23}.

Finally, the heating of the IGM before reionization is governed by X-ray 
emission from the first galaxies. It is likely that the X-ray 
luminosities from the first galaxies were dominated by high mass 
X-ray binary stars (HMXBs; e.g., \citealt{fragos13,pacucci14}). 
The X-ray luminosities of the HMXB population depends on the 
star formation rate and metallicity of the galaxy. We thus sampled 
the X-ray luminosities of the galaxy population from another 
log-normal distribution, whose mean $L_X/\textrm{SFR}$ was a 
double power law dependent on SFR and stellar mass, 
via a gas-phase metallicity $Z$,
\begin{equation}\label{eq:xraymean}
    \mu_{\mathrm{X}} = \frac{\mathrm{SFR}}{M_\odot \mathrm{yr}^{-1}} \times L_{X,\mathrm{norm}} \left( \left( \frac{Z}{0.05 Z_\odot} \right)^{0.64} + 1 \right)^{-1}
\end{equation}
\noindent where the fiducial choices were motivated by local HMXB 
luminosity functions (e.g., \citealt{lehmer19}). We computed the 
metallicity using the redshift-adjusted fundamental mass-metallicity  
relation (FMZR) from  \citet{curti2020},
\begin{equation}
    \frac{Z}{Z_\odot} = 1.23 \left( 1 + \left( \frac{M_*}{M_0} \right)^{-2.1} \right)^{-0.148}
    10^{-0.056z + 0.064} ~, 
\end{equation}
\noindent where
\begin{equation}
    M_0 = 10^{10.11} \left( \frac{\mathrm{SFR}}{M_\odot yr^{-1}} \right) ^{0.56}.
\end{equation}

This flexible galaxy model is anchored in well-established empirical 
relations (e.g., SHMR, SFMS, and FMZR).  The free parameters of these 
relations, including the means and scatters, might eventually be 
inferred from observations. However, because we provide 
forecasts for upcoming surveys, we chose fiducial values motivated 
by a combination of observations and hydrodynamic simulations. 
Table \ref{tab:galax_params} summarizes the 
fiducial values for all free parameters (for a further 
motivation of these fiducial choices, see \cite{nikolic24} and \cite{davies25}).

With the above galaxy model defining UV and X-ray emissivities, 
\texttt{21cmFAST} computed the associated inhomogeneous cosmic 
radiation fields by tracking the evolution of the temperature and 
ionization state in each simulation cell (for more details, see 
\citealt{mesinger07,mesinger2011,davies25}).

The 21-cm brightness temperature of each cell was then computed as (e.g., \citealt{furlanetto04})

\begin{eqnarray}
\nonumber   \delta T_b\approx 27\text{mK}
   \left(\frac{\Omega_{b,0}h^2}{0.023}\right)
   \left(\frac{0.15}{\Omega_{m,0}h^2}\frac{1+z}{10}\right)^{1/2}
   \\ \times x_{\text{HI}}\Delta
   \left(1-\frac{T_R}{T_S}\right)
   \left(\frac{H}{\partial_r v_r}\right),
\end{eqnarray}
\noindent where $x_\text{HI}$ is the neutral hydrogen fraction, 
$\Delta \equiv \rho/\bar{\rho}$ is the overdensity, $T_R$ is the CMB temperature, $T_S$ is the spin 
temperature, and $\partial_r v_r$ is line-of-sight gradient of the 
gas velocity. 

The upper row of Figure \ref{fig:cutout} shows 2D slices through the 21-cm 
brightness temperature field in our fiducial model at redshifts $6$, $7$, and $8$. 
We also show a 200 Mpc on a side zoom-in in the bottom row, together with the corresponding 
galaxy field.  The $M_{\rm UV}< -17$ galaxies shown in the bottom row reside in 
ionized regions with about zero 21-cm emission/absorption.  Our fiducial parameter choices 
result in a midpoint (end) of reionization at $z=7.7$ ($5.5$), which is consistent with the latest 
observational estimates (e.g., \citealt{qin24}). Prior to the EoR, the 21cm signal on large scales is driven by temperature fluctuations, because high mass X-ray binaries (HMXBs) inside the first galaxies are expected to heat the IGM to temperatures above the CMB temperature (e.g., \cite{furlanetto06}).  This marks the so-called Epoch of Heating (EoH; e.g. \cite{mesinger14}), when the 21-cm signal changed from being seen in absorption to being seen in emission against the CMB. This also marks a sign change in the galaxy--21cm cross-power (e.g., \citealt{heneka20,hutter23,moriwaki24}; see Fig. \ref{fig:ps1d} and associated discussion).

\subsection{Galaxy selection based on nebular lines}

\texttt{21cmFASTv4} produced a catalog of galaxies through 
sourcing cosmic radiation fields, which we then postprocessed 
with observational selection criteria. 
Every mock survey we considered required 
its own galaxy map, biased by the corresponding 
selection criteria and instrument sensitivity.

Unfortunately, selection through broadband 
photometry results in redshift uncertainties 
that are too large to estimate a cross-power 
spectrum with 21-cm (e.g., \citealt{lidz09,la_plante23}). 
We therefore based our selection criteria on nebular
emission lines. These can be used for narrow-band dropout surveys 
(e.g., Ly-$\alpha$ emitters), low-resolution image spectroscopy (e.g., grism/prism), 
and high-resolution slit-spectroscopy (requiring follow-up of a 
photometric candidate sample). We quantify each of these scenarios below. 

We considered two fiducial selection criteria: one targeting 
Lyman-$\alpha$, and another which selects galaxies on the basis 
of H$\alpha$/H$\beta$ and [OIII] emission. 
While Lyman-$\alpha$ is generally intrinsically 
brighter than other nebular lines, it is more 
sensitive to ISM/CGM/IGM absorption. 
Therefore, the observed Ly$\alpha$ luminosities 
have a large sightline-to-sightline variability.

Below, we discuss our emission line models. 
These were shown to reproduce various observables, 
including the rest-frame UV and Lyman-$\alpha$ luminosity 
functions as well as Lyman-$\alpha$ equivalent width distributions (e.g., 
\citealt{kennicut98, ly07, lim17, mason2018, mirocha2017, park2019}).

\subsubsection{Lyman alpha}

We related the emergent (escaping the galaxy after passing through 
the ISM and CGM) Lyman-$\alpha$ luminosity to the UV magnitude 
of the galaxy (see Equation \ref{eq:mab}) following 
\citet{mason2018}. The rest-frame, emergent Ly-$\alpha$ 
equivalent width (EW) is defined as
\begin{equation}
    \label{eq:ew}
    W=\frac{1216\AA}{2.47\cdot10^{15}\textrm{Hz}}\left(\frac{1500}{1216}\right)^{\beta+2}10^{0.4(M_{UV}-51.6)}L_{\textrm{Ly}\alpha}^{\textrm{emerg}}
\end{equation}

\noindent where $L_{\textrm{Ly}\alpha}^{\textrm{emerg}}$ 
is the emergent luminosity of the Ly-$\alpha$ line and 
$\beta$ is the slope of the UV continuum emission, which we took to be
\begin{equation}
    \beta = -2.05\pm0.16-(0.2\pm0.07)\times(M_{UV}+19.5)
\end{equation}
\noindent following \citet{bouwens2014}.
The probability of a galaxy having a nonzero emergent 
Ly-$\alpha$ emission is
\begin{equation}
    \label{eq:a}
    A(M_{UV}) = 0.65 + 0.1 \tanh{[3(M_{UV}+20.75)]}.
\end{equation}
\noindent and the EWs of Ly-$\alpha$ 
emitters follow the distribution 
\begin{equation}
    \label{eq:prob}
    P(W) \propto \exp{\left[-\frac{W}{W_c(M_{UV})}\right]},
\end{equation}
\noindent where $W_c$ is the characteristic Ly-$\alpha$ 
equivalent width, itself a function of $M_{UV}$,
\begin{equation}
    \label{eq:wc}
    W_c(M_{UV}) = 31 + 12 \tanh{[4(M_{UV}+20.25)]}.
\end{equation}

\noindent In mapping from $M_{\rm UV}$ to $L_{\rm Ly\alpha}^{\rm emerg}$, we drew the probability of emission 
and the equivalent width of each galaxy from the above distributions and then solved for the luminosity via Equation 
\ref{eq:ew}.  \cite{mason2018} determined the parameters of Equations \ref{eq:a} and \ref{eq:wc} by fitting $\tanh$ 
functions to post-EoR ($5\lesssim z \lesssim 6$) galaxies from a large program using VLT/FORS2 as well as the VANDELS survey 
\citep{debarros17, talia23}. 

We assumed that all flux blueward of the circular 
velocity of the host halo is scattered out of the 
line of sight by the CGM. To account for IGM attenuation, we multiplied 
$L_{\textrm{Ly}\alpha}^{\rm emerg}$ by  $\exp\{-\tau_{\textrm{EoR}}\}$, 
where $\tau_{\textrm{EoR}}(\lambda)$ is the cumulative optical depth of 
cosmic HI patches along the line of sight to the galaxy. 
We sampled $\tau_{\textrm{EoR}}$ from a conditional 
log-normal distribution motivated by 
\cite{mesinger2008}, whose parameters depend on the mean 
IGM neutral fraction and the mass of the halo hosting the galaxy 
(for more details on this method, we refer to Appendix \ref{apdx:tau}).

When determining whether a particular galaxy was detected in 
Lyman-$\alpha$, we compared the observed flux (i.e. 
emergent flux attenuated by the IGM) with the flux limit of the instrument of interest. For example, a  follow-up of a 
$M_{\textrm{UV}}\leq-20$ 
photometric candidate at $z=6$, using a spectrograph with a 5$\sigma$ AB magnitude limit of $m_{\rm AB}^\alpha \leq 26$ targeting Lyman-$\alpha$ can only 
detect Ly-$\alpha$ if the observed equivalent width is greater than $\geq8$ $\AA$. 

\begin{figure*}
    \resizebox{\hsize}{!}
    {\includegraphics[width=0.45 \textwidth]{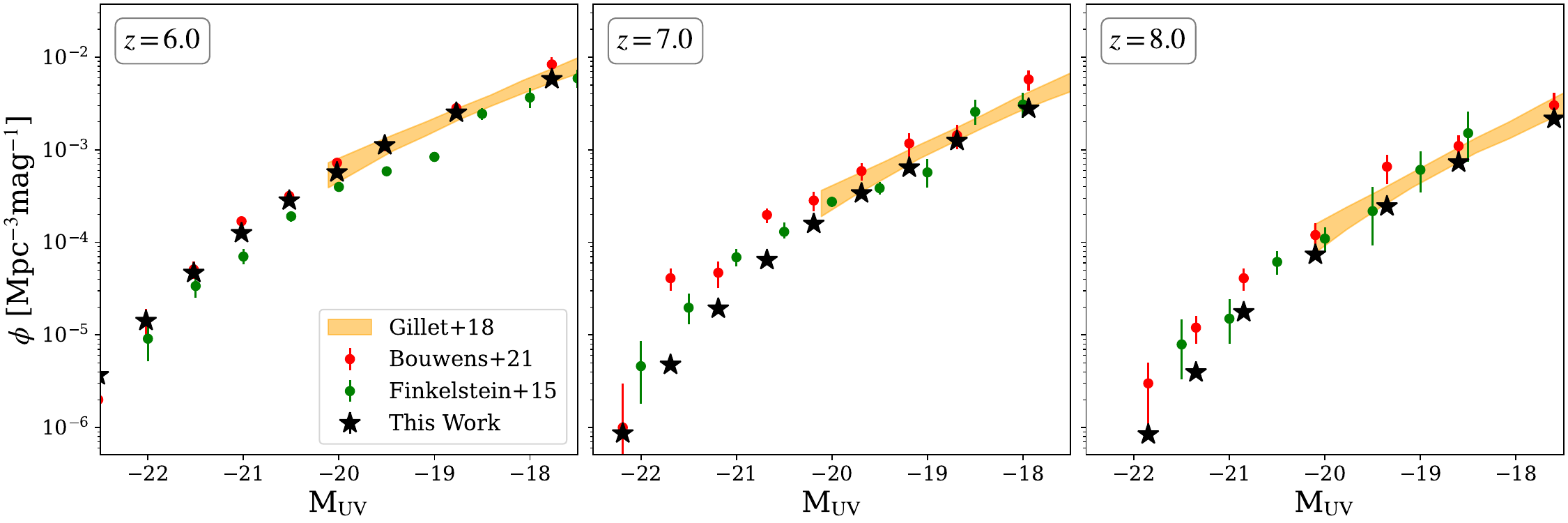}}
    \caption{UV luminosity functions (LFs) at $6\leq z\leq 8$ from our fiducial model, 
    compared with several observational estimates. We show {\it Hubble} estimates from \cite{bouwens21}
    and \cite{finkelstein15}, as well as the observation-averaged posterior 
    from \cite{gillet20} ($1\sigma$ credible interval).
 }
    \label{fig:LF}
\end{figure*}

\begin{figure}
    \resizebox{\hsize}{!}
    {\includegraphics[width=0.45 \textwidth]{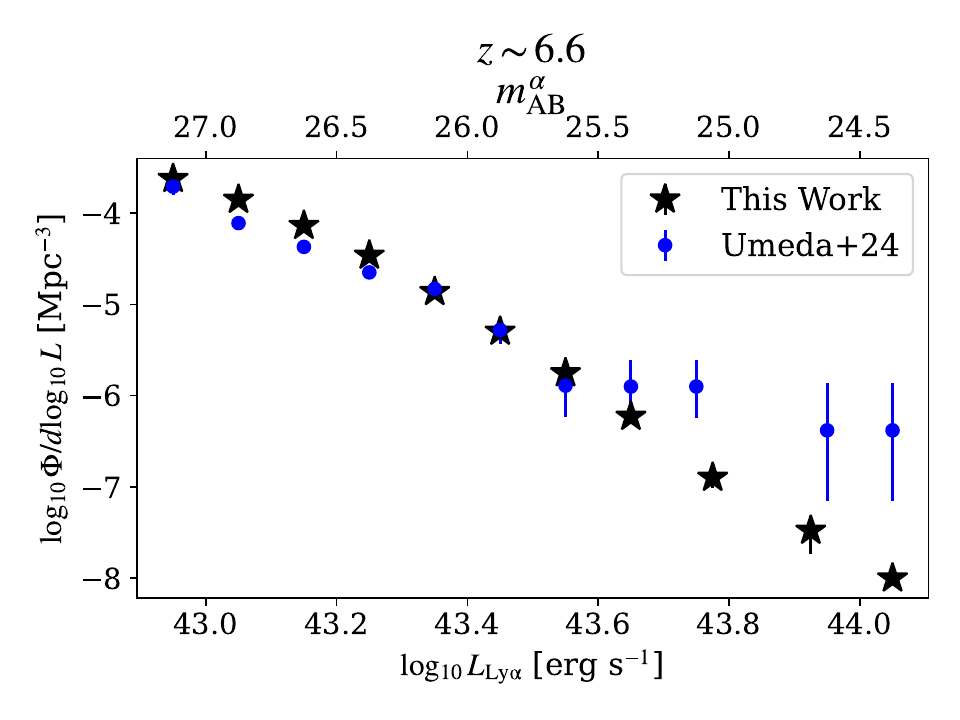}}
    \caption{Lyman-$\alpha$ luminosity functions at $z=6.6$.  Our fiducial model is 
    based on equivalent width distributions of photometrically selected 
    candidates (see text for details). We also show narrow-band selected 
    LAE LFs from Subaru \citep{umeda24}.
 }
    \label{fig:LAELF}
\end{figure}

Figure \ref{fig:LF} shows UV luminosity functions (LFs) of our fiducial model
at $z=6$, $7$, and $8$ (left to right). The simulated LFs agree with 
observational estimates (we show some examples from 
\citealt{finkelstein15, bouwens21, gillet20}). 
In Figure \ref{fig:LAELF} we also show the corresponding 
Lyman-$\alpha$ LFs at $z\sim6.6$, together with narrow-band 
selected Lyman-$\alpha$ emitter (LAE) LFs from the Subaru 
telescope \citep{umeda24}. Our fiducial model agrees reasonably well 
with the observational estimates (within 2$\sigma$), 
even though it uses empirical relations from 
photometrically selected galaxies. There is some evidence that our 
LAE LFs are lower than observations at the bright end; if this were 
confirmed, it would make our fiducial estimates of the cross-power 
spectrum based on Lyman-$\alpha$ conservatively low.

\subsubsection{Selection with other lines}

We also considered spectroscopic confirmation targeting other 
nebular lines. We used the planned JWST survey as a template: the First 
Reionization Epoch Spectroscopically Complete Observations 
(FRESCO; \citealt{fresco}). FRESCO will target a 
field of $62$ square arcminutes at a depth of 
$m_{\textrm{AB}}\sim28.2$ using the NIRCam/grism instrument. 
NIRCam filters are sensitive to H$\alpha$ below redshift $7$ 
and sensitive to [OIII] and H$\beta$ for $z\in(7,9)$.

We produced separate galaxy maps using the above 
line selection criterion for various magnitude cutoffs in 
order to forecast prospects for the 21-cm cross-correlation with 
FRESCO or a FRESCO-like field. We made the common assumption 
that the luminosity of these low-opacity nebular lines 
traces the star formation rate of the galaxy 
(e.g., \citealt{kennicut98, ly07, lim17}),
\begin{equation}
    L_{\textrm{H}\alpha}=\textrm{SFR}\times1.27\cdot10^{41}\frac{\textrm{erg}\cdot\textrm{yr}}{\textrm{s}\cdot\textrm{M}_\odot},
\end{equation}
\begin{equation}
    L_{\textrm{[OIII]}}=\textrm{SFR}\times1.32\cdot10^{41}\frac{\textrm{erg}\cdot\textrm{yr}}{\textrm{s}\cdot\textrm{M}_\odot},
\end{equation}
\noindent and
\begin{equation}
    L_{\textrm{H}\beta}=\textrm{SFR}\times7.14\cdot10^{40}\frac{\textrm{erg}\cdot\textrm{yr}}{\textrm{s}\cdot\textrm{M}_\odot},
\end{equation}
\noindent where the SFR is in units of M$_\odot$/yr and the luminosity is in erg/s.
\begin{table}
    \centering
    \caption{Fiducial galaxy parameters.}
    \begin{tabular}{c|c|c}
        \hline
         Parameter & Description & Value \\
         \hline
         $f_{*,10}$ & \tiny SHMR normalization & 0.05 \\
         $\alpha_*$ & \tiny Power-law index of SHMR & 0.5 \\
         $\alpha_{*2}$ & \tiny High mass power-law index of SHMR & -0.61 \\
         $M_\mathrm{p1}$ & \tiny lower pivot mass in SHMR & $10^{10} M_\odot$ \\
         $M_\mathrm{p2}$ & \tiny upper pivot mass in SHMR & $2.8 \times 10^{11} M_\odot$ \\
         $\sigma_*$ & \tiny lognormal scatter in SHMR & 0.3 dex \\
         $C_*$ & \tiny auto-correlation timescale of SHMR & 0.5 \\
         $t_*$ & \tiny SSFR normalization & 0.5 \\
         $\sigma_\mathrm{SFR,lim}$ & \tiny minimum lognormal scatter in SSFR & 0.19 dex \\
         $\sigma_\mathrm{SFR,idx}$ & \tiny Power-law scaling of SSFR scatter & -0.12 \\
         $C_\mathrm{SFR}$ & \tiny auto-correlation timescale of SSFR & 0.1 \\
         $L_{X,\mathrm{norm}}$ & \tiny $L_X / \mathrm{SFR}$ normalization & $10^{40.5} \mathrm{erg s}^{-1}$ \\
         $\sigma_X$ & \tiny lognormal scatter in $L_X / \mathrm{SFR}$ & 0.5 dex \\
         $C_\mathrm{X}$ & \tiny auto-correlation timescale of $L_X / \mathrm{SFR}$ & 0.5 \\
         $f_\mathrm{esc,10}$ & \tiny Escape fraction normalization & 0.1 \\
         $\alpha_\mathrm{esc}$ & \tiny Escape fraction power-law scaling & 0.5\\
         \hline
    \end{tabular}
    \label{tab:galax_params}
\end{table}

\section{Computing the S/N of the cross-power spectrum}
\label{sec:det}
\begin{figure}
    \resizebox{\hsize}{!}
    {\includegraphics[width=\textwidth]{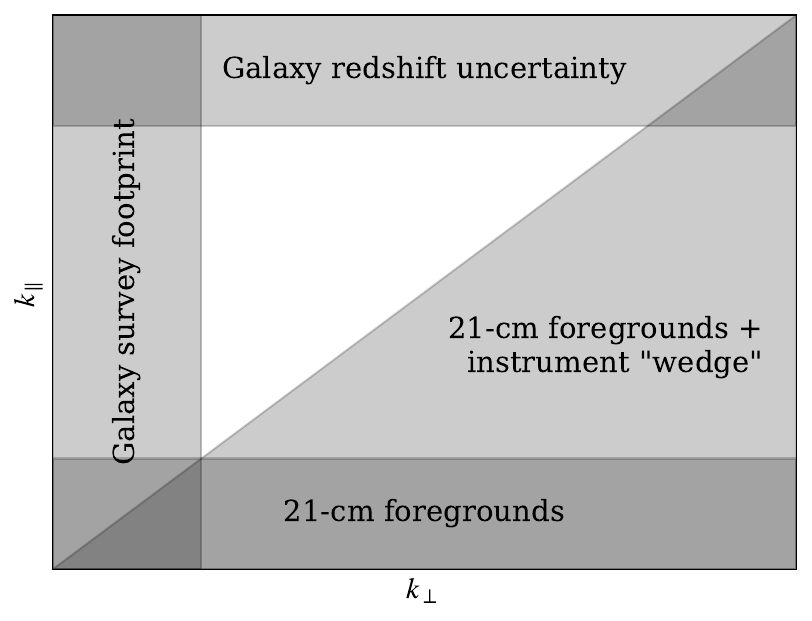}}
    \caption{Schematic illustrating  regions in the cylindrical 
    power spectrum that contribute negligibly to the
    cross-power S/N due to instrument, survey, or systematic 
    limitations.}
    \label{fig:toy}
 \end{figure}

After computing the cosmic signals, 
we determined the detectability of 
the galaxy--21-cm cross-power spectrum given a specific
pair of instruments. We computed the power 
spectrum in cylindrical coordinates $(k_\perp,k_\parallel)$ 
because this is the natural basis for interferometers that seek to
maximize the S/N of a detection. Furthermore, this
basis relegates foreground contamination to an easily excised 
(at least in principle) corner of Fourier space.
Here, $k_\perp$ refers to wavemodes in the sky plane, 
and $k_\parallel$ refers to wavemodes along the line of 
sight. Thus, a cylindrical cross-power spectrum was 
constructed by independently averaging Fourier modes 
along the line of sight and in sky-plane bins. 
 
We split each lightcone into 
redshift chunks, each with a frequency-domain depth of $\Delta \nu=6$ MHz centered on redshift $z$. 
From each chunk, we computed the cross-power spectra,
\begin{equation}
    P_{21,g}(k_\perp,k_\parallel,z)
    =\langle\delta T^*_b(z)\delta_g(z)\rangle_{k_\perp,k_\parallel},
\end{equation}
\noindent where $P_{21,g}(z)$ is the galaxy--21-cm 
cross-power spectrum, $\delta T_b(z)$ is the 
21-cm brightness temperature field, 
and $\delta_g(z)\equiv n_g(z)/\bar{n}_g(z)-1$ is the 
galaxy overdensity field.

The averaging in the above equation was 
computed only over wavemodes that are accessible to a specific 
instrument or survey and that were not highly contaminated by 
foregrounds or systematics. Averaging over systematics-contaminated 
modes might still improve the S/N, because that uncorrelated 
systematics do not impact the mean of the cross-power, 
even though they increase the variance (e.g., \cite{fronenberg24}). 
In order to quantify exactly which wavemodes should be excised, 
however, we would need to forward-model systematics or foreground 
residuals. We save this for future work and made the simplifying 
assumption that foreground- or systematics-dominated wavemodes are unusable while the remaining wavemodes include no residual systematics.

We show a schematic of these contaminated or missing regions in 2D Fourier space 
in  Figure \ref{fig:toy}.  Each shaded region illustrates some range of 
wavemodes that contribute negligibly to the cross-power due 
to some properties of the instrument/survey/systematics. These include the area of the overlap 
between the galaxy field and 21-cm field (limited by the galaxy survey 
footprint), the redshift uncertainty of the galaxies, and contamination 
by 21-cm foregrounds. 

The galaxy survey will typically
have the smaller of the two footprints
\footnote{For example, the wedding-cake strategy envisioned for the 
SKA-low Key Science Project is expected to cover up to $\sim10^4$ sq. 
deg., while the HERA observing stripe covers $\sim4000$ sq. deg.}, and 
it can range from tens to hundreds of square degrees for narrow-band 
dropout or grism surveys to $\lesssim$ sq. deg for deep follow-up of 
photometric candidates using high-resolution spectroscopy. The area of 
the galaxy survey sets the largest on-sky scale 
(smallest $k_\perp$) accessible for computing the cross-power 
spectrum. Similarly, the galaxy redshift uncertainty effectively sets 
the smallest accessible line-of-sight scale (largest $k_\parallel$; as 
described further below, we did not excise the larger $k_\parallel$ 
modes, but instead accounted for the corresponding uncertainty when we 
computed the noise). The redshift uncertainty we considered 
ranged from $\delta z \sim 0.05$ for narrow-band LAE surveys to 
$\delta z \sim 0.001$ for spectroscopy\footnote{In principle, the 
redshift uncertainty is determined by a combination of instrument 
resolution, peculiar velocity, and the offset of the Lyman-$\alpha$ 
line from systemic. The latter two effects set a lower limit for the 
achievable redshift uncertainty, which is about $\delta z \sim0.001$. 
We postpone a more careful treatment of each of these terms to future 
work focusing on specific instruments.} (see Table \ref{tab:experiments}).

Finally, the spectral smoothness of 21-cm foregrounds excludes 
the lowest $k_\parallel$ scales. More importantly, 
the response function of 21-cm interferometers results in 
leakage of foreground emission into a wedge-shaped region 
of Fourier space, that overwhelms the cosmic signal 
(e.g., \citealt{Datta2010,Pober2014a,Dillon2014,liu14a,Pober2014b}). 
To quantify the wedge-contaminated region, we used different slopes in cylindrical $k$-space (e.g., \citealt{thyagarajan15}), 
\begin{equation}
    m(z)\equiv\frac{k_\parallel}{k_\perp}=\sin(\theta_{\text{FoV}})
    \frac{D_c(z)H(z)}{c(1+z)},
    \label{eq:wedge}
\end{equation}
\noindent where $D_c(z)$ is the comoving distance to that redshift, and 
$\theta_{\mathrm{FoV}}$ is the angular radius of the field of view of 
the beam of the 21-cm interferometer (see also \citealt{munshi25} for an 
alternative characterization of the wedge). We considered three 
foreground scenarios in this study: a fiducial scenario in which
only $k$-modes up to the slope in Equation \ref{eq:wedge} 
are foreground-dominated, a pessimistic scenario in which we 
doubled the slope, and an optimisic scenario where we halved the 
slope to represent the recovery of lost Fourier modes via advanced 
foreground-mitigation techniques
(e.g., \citealt{Hothi_2020, sgh21, bianco24, kennedy24}). Foreground 
scenarios typically vary from horizon-limited to beam-limited, with the 
former corresponding to our fiducial scenario and the latter to our 
pessimistic scenario. We also varied the extent of intrinsic foregrounds 
in each scenario by excising $k_\parallel\leq0.07$ Mpc$^{-1}$ for the 
fiducial and optimistic scenarios and $k_\parallel\leq0.175$ Mpc$^{-1}$ 
for the pessimistic scenario \citep{kubota2018}.

After we defined the cross-power spectrum as well as the region in 
cylindrical Fourier space over which it was computed, we estimated the 
associated uncertainty. The variance of the cross-power spectrum is 
(see \citealt{lidz09})
\begin{equation}
    \sigma^2_{21,g}(k_\perp,k_\parallel)=\text{var}\left[\frac{1}{T_0(z)}P_{21,g}(k_\perp,k_\parallel)\right],
\end{equation}
\noindent which reduces to
\begin{equation}
    \sigma^2_{21,g}(k_\perp,k_\parallel)=\frac{1}{2}\left[P^2_{21,g}(k_\perp,k_\parallel)+\sigma_{21}(k_\perp,k_\parallel)\sigma_g(k_\perp,k_\parallel)\right].
\end{equation}
\noindent Here, $T_0$ is a dimensionless normalization factor defined as

\begin{equation}
    T_0(z)=26\left(\frac{T_S-T_\gamma}{T_S}\right)\left(\frac{\Omega_b h^2}{0.022}\right)
    \left[\left(\frac{0.143}{\Omega_m h^2}\right)\left(\frac{1+z}{10}\right)\right]^{1/2},
\end{equation}

\noindent where $z$ is the redshift, $T_S$ is the 21-cm spin 
temperature, $T_\gamma$ is the CMB temperature, $\Omega_b$ and 
$\Omega_m$ are the baryonic and total matter densities of the 
Universe, respectively, and $h$ is the dimensionless Hubble parameter.

Similarly to the cross-power, the variance of the auto-power 
spectra can be written as a sum in quadrature of the cosmic 
variance and noise power,
\begin{equation}
    \sigma^2_{21}(k_\perp,k_\parallel)=\frac{1}{2}\left[\frac{1}{T^2_0(z)}P_{21}(k_\perp,k_\parallel)+P_{21}^{\text{noise}}(k_\perp,k_\parallel)\right]^2
\end{equation}
\noindent and
\begin{equation}
    \sigma^2_{g}(k_\perp,k_\parallel)=\frac{1}{2}\left[P_{g}(k_\perp,k_\parallel)+P_{g}^{\text{noise}}(k_\perp,k_\parallel)\right]^2,
\end{equation}
\noindent where the first terms on the right sides correspond to 
cosmic variance, while $P^{\text{noise}}_{21}$ and 
$P^{\text{noise}}_{g}$ correspond to the 21-cm thermal noise and the 
galaxy redshift uncertainty, respectively.  We describe these in turn 
below.

\begin{figure}
    \resizebox{\hsize}{!}
    {\includegraphics[width=\textwidth]
    {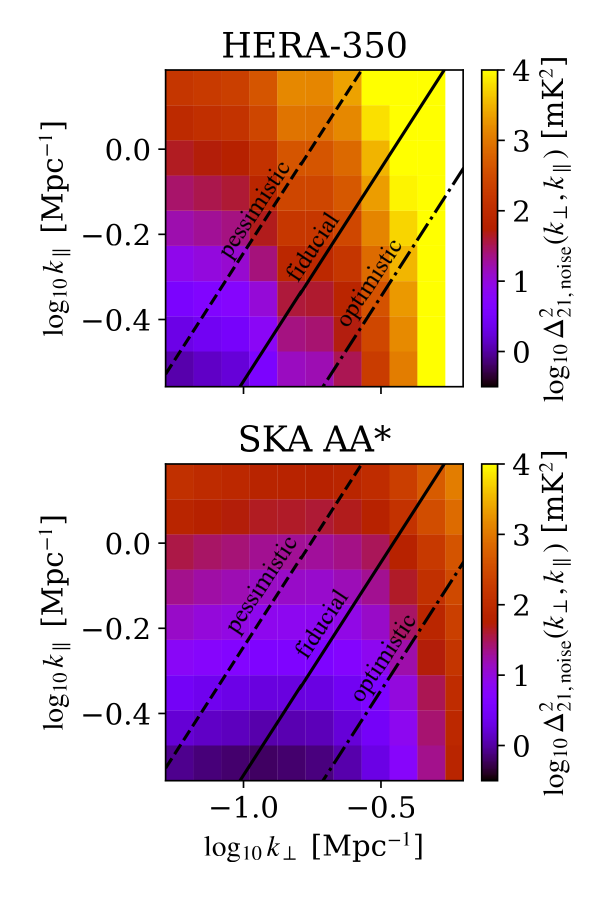}}
    \caption{The cylindrical noise power spectra of HERA-350 ({\it top})
    and SKA-low AA* ({\it bottom}) for a fiducial 
    $1000$ hour observation at redshift $6.1$. The diagonal lines in each 
    panel denote the three choices for wedge excision considered in this work.}
    \label{fig:ps_noise}
\end{figure}

We computed the 21-cm thermal noise power spectrum, 
$P^{\textrm{noise}}_{21}$, using \texttt{21cmSense} 
\citep{Pober2014a,21cmsense}. We used antenna layouts corresponding 
to the full HERA-350 array \citep{hera24}, as well as the initial 
AA* deployment of SKA-low \citep{skao23}. We assumed a $1080$h total integration of drift scan for HERA-350/SKA-low 
with $6$ hours of observation per night over $180$ days
\footnote{The SKA-low AA* beam covers a stripe on the sky with a 
width of $3.68^\circ$. Integrating for $6$ hours per night over the 
same stripe produces a survey footprint of $332$ sq. deg.. 
Similarly, HERA covers a stripe of width $11.0^\circ$ and thus produces a
survey footprint of $990$ sq. deg. \citep{21cmsense}. We assume 
Gaussian beams, and the widths quoted above correspond to the 
$3\sigma$ extents.}.
The resulting thermal noise is plotted in Figure \ref{fig:ps_noise}. 
Our mock survey with SKA-low resulted in overall lower noise. 
This is most evident at the largest $k_\perp$, which are better sampled 
because the SKA-low has a greater number of long core baselines. 
We also show the three different 
wedge scenarios we considered with diagonal lines. 
Optimistic foreground removal would benefit SKA-low 
far more than HERA-350, since these wedge modes in HERA have 
very high levels of thermal noise. Foreground-avoidance 
was indeed a design choice for HERA. 

We adopted the galaxy noise model of \cite{lidz09}, wherein the shot noise sets the minimum noise level and the redshift uncertainty governs the line-of-sight positional uncertainty,
\begin{equation}
    P^{\text{noise}}_g(k_\parallel)=\frac{1}{n_{\text{gal}}}\exp{\left[k_\parallel^2\left(\frac{c\sigma_z}{H(z)}\right)^2\right]},
    \label{eq:pg}
\end{equation}
\noindent where $n_{\text{gal}}$ is the number density of galaxies in the survey volume and $\sigma_z$ is the redshift uncertainty set by the instrument. This form for $P_g^{\rm noise}(k_\parallel)$ states that the galaxy power spectrum is suppressed by a Gaussian kernel whose width equals the characteristic scale of redshift uncertainties in the field \citep{seo03}. \cite{fisher93} and \cite{feldman1994} showed that this suppression arises naturally in measured galaxy power spectra. As shown in Table \ref{tab:experiments}, we used three fiducial values for $\sigma_z$ that corresponded to narrow-band dropouts (LAE; e.g. \citealt{silverrush2018}), imaging (i.e., slitless) spectroscopy (grism/prism; e.g. \citealt{hls2022,fresco}), and slit spectroscopy (e.g., \citealt{mosaic, moonrise}).

The signal-to-noise ratio in each bin $(k_\perp,k_\parallel)$  depends not only on the cross-power amplitude and uncertainty, but also on the number of Fourier modes sampled by that bin (i.e., sample variance). When the bins are of equal length in log-space, the number of modes per bin is
\begin{equation}
    dN(k_\perp,k_\parallel)=\frac{k_\perp^2k_\parallel V_{\text{survey}}}{(2\pi^2)}d\ln k_\perp d\ln k_\parallel,
    \label{eq:modes}
\end{equation}
\noindent where $V_{\text{survey}}$ is the volume of the survey in comoving units. The signal-to-noise ratio in a single bin adds in quadrature with the number of Fourier modes sampling that bin:
\begin{equation}
    \hat{s}(k_\perp,k_\parallel)=\sqrt{dN(k_\perp,k_\parallel)}\frac{P_{21,g}(k_\perp,k_\parallel)}{\sigma_{21,g}(k_\perp,k_\parallel)}.
    \label{eq:ps_snr}
\end{equation}
Following \cite{furlanetto07}, we 
computed the global signal-to-noise ratio as the sum in 
quadrature of\footnote{Our choice to treat 
the global S/N as the quadrature sum 
of the S/N in each bin reflects our assumption that 
the noise in each Fourier mode is statistically independent.
This assumption results in a modest overestimation of 
the signal-to-noise ratio (e.g., \cite{liu14a}, \cite{liu14b}, \cite{prelogovic23}).} 
the signal-to-noise ratios in each $(k_\perp,k_\parallel)$ bin and $z$,
\begin{equation}
    \text{S/N}^2=\sum_{k_\perp,k_\parallel,z}\hat{s}^2(k_\perp,k_\parallel,z).
    \label{eq:snr}
\end{equation}
\begin{figure*}
    \resizebox{\hsize}{!}
    {\includegraphics[width=\textwidth]
    {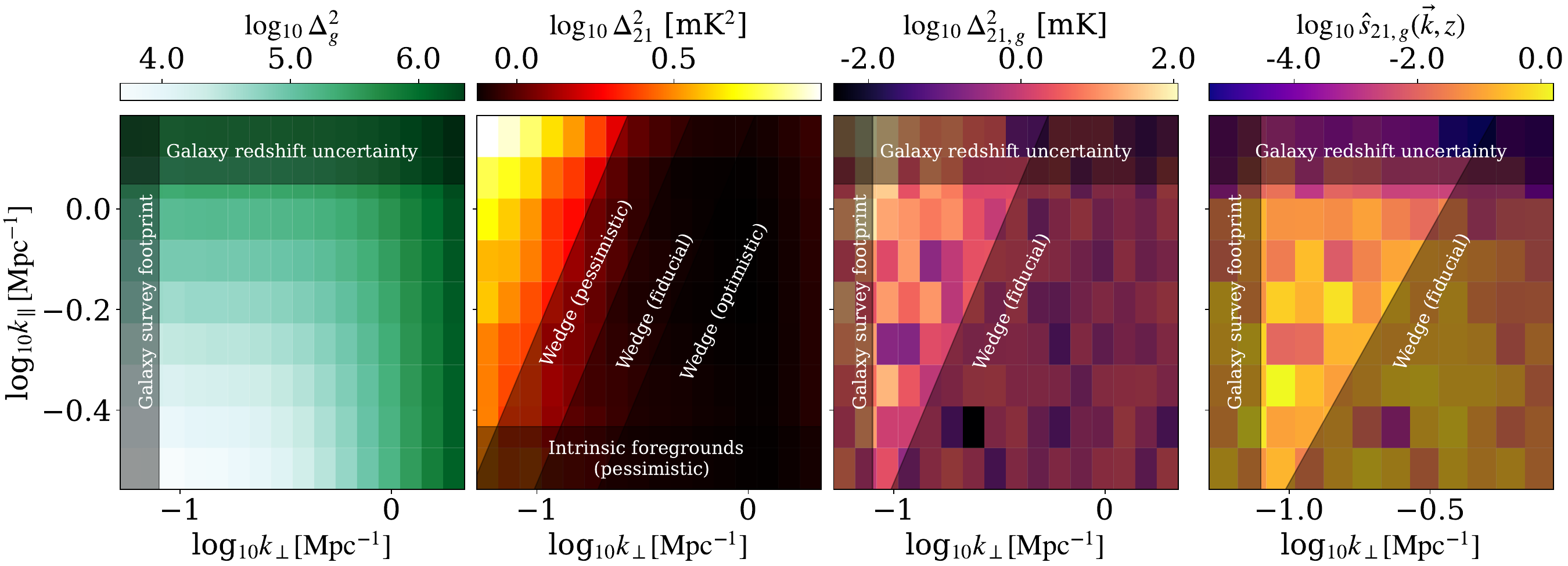}}
    \caption{Power spectra of the galaxy field, 21-cm brightness temperature field, their cross-power spectrum,
    and the signal-to-noise ratio of the cross-power spectrum as a function of $k_\parallel$ and $k_\perp$ for a
    fiducial scenario (left to right). Overlaid are various noise-dominated/excluded regions imposed
    by systematics and observational effects. The 21-cm brightness temperature power 
    spectrum includes cuts showing the three foreground+instrument (i.e. 
    wedge) contamination scenarios we considered. In this illustrative example, the galaxy survey footprint corresponds to $\sim1$ sq. deg. at $z\sim6$, and the galaxy redshift uncertainty corresponds to $\delta z \sim 0.01$, which is characteristic of slitless spectroscopy.}
    \label{fig:megaplot}
 \end{figure*}

Figure \ref{fig:megaplot} illustrates the principal data products in our pipeline for estimating the signal-to-noise 
ratio of the cross-power spectrum. The first two panels from the left show the galaxy and 21-cm auto-power spectra, 
respectively, as computed from a 1 Gpc coeval cube at $z=6.1$. The third panel shows the galaxy--21-cm cross-power spectrum of the same cube. The correlation between the $k$-modes is weaker than in the galaxy or 21-cm spectra. 
The rightmost panel shows the resulting signal-to-noise ratio 
from Equation \ref{eq:ps_snr}. Superimposed on each panel are shaded 
regions corresponding to wavemodes that are inaccessible due to 
survey/instrument/systematics limitations. The final S/N 
primarily depends on the unshaded wavemodes in the rightmost panel, 
which are extended to also include other redshift bins.

\section{Mock observations}
\label{sec:exp}

\begin{table*}
    \centering
    \caption{21-cm and galaxy survey parameters considered.}
    \begin{tabular}{c|c|c|c|c|c|c}
        \hline
        \hline
         \textbf{Interferometer} & \textbf{Detection Type} & \textbf{Line Targeted} & $\bm{\delta}\mathbf{z}$ & \textbf{Footprint Range} & \textbf{Depth Range} ($\bm{m}_{\textbf{AB}}$)  & \textbf{Figure(s)} \\
         \hline
         SKA-low & Narrow-band dropout & Ly-$\alpha$ & $5\cdot10^{-2}$ & $10-20$ sq. deg. & $25.5-30$ & 8,9,10 \\
         SKA-low & Grism $z>7.2$ & Ly-$\alpha$ & $10^{-2}$ & $0-25$ sq. deg. & $25-27$ & 8,9,10  \\
         SKA-low & Spectroscopy & Ly-$\alpha$ & $10^{-3}$ & $0-4$ sq. deg. & $24-27$  & 8,9,10 \\
         HERA & Narrow-band dropout & Ly-$\alpha$ & $5\cdot10^{-2}$ & $10-20$ sq. deg. & $25.5-30$  & 11 \\
         HERA & Grism $z>7.2$ & Ly-$\alpha$ & $10^{-2}$ & $0-120$ sq. deg. & $25-27$  & 11  \\
         HERA & Spectroscopy & Ly-$\alpha$ & $10^{-3}$ & $0-4$ sq. deg. & $24-27$  & 11  \\
         SKA-low & Grism & H$\alpha$/[OIII] & $10^{-2}$ & $0-350$ sq.arcmin. & $27-30$  & 12 \\
         \hline
         \hline
    \end{tabular}
    \label{tab:experiments}
\end{table*}

The galaxy--21-cm cross-power depends on the telescopes used for 
each observation as well as on the survey strategies. 
As discussed above, our fiducial 21-cm observations are $1080$h 
total integration with either HERA-350 or SKA-low AA*. 
For simplicity, we did not vary the 21-cm survey specifications.

We considered three general types of instruments that are used 
for galaxy surveys, in order of increasing redshift accuracy
\footnote{As mentioned above, we do not consider broad-band photometry (i.e. Lyman break galaxy surveys), because the corresponding redshift 
uncertainties are too large to be useful for cross correlation with 21-cm (e.g., \citealt{la_plante23}).}:
\begin{enumerate}
\item {\it Narrow-band dropout} uses narrow photometric bands to identify Lyman-$\alpha$ line emission. 
Galaxies thus identified are commonly referred to as LAEs, and their 
associated redshift uncertainty is determined by the photometric bands used in the drop-out selection. We took Subaru as a fiducial 
telescope, assuming a redshift uncertainty of 5\% \citep{silverrush2018}.
\item {\it Imaging (slitless) spectroscopy} by slitless spectrographs (e.g., grism) provides low-resolution multiple spectra within a contiguous field, without requiring a preselection of candidates through broad-band photometry. We took {\it Roman} as our 
fiducial telescope for LAE-selected surveys and {\it JWST} (NIRCam) for H$\alpha$/H$\beta$/OIII-selected surveys. We assumed a redshift uncertainty of 1\% for the two instruments \citep{hls2022}.
\item {\it Slit spectroscopy} of high resolution spectrographs result in the smallest redshift uncertainty. Slit spectroscopy requires follow-up of photometric candidates, however, either from existing fields or from new fields. We assumed that photometric candidates of sufficient depth are available (e.g., Euclid Deep Field South \citealt{euclid24}, SILVERRUSH \citealt{silverrush2018}, legacy {\it Hubble} fields), and the bottleneck comes from the spectroscopic follow-up of Lyman-$\alpha$. We adopted a slit spectroscopy redshift uncertainty of $0.1$\% (e.g., \citealt{moonrise}).
\end{enumerate}

\begin{figure}
    \resizebox{\hsize}{!}
    {\includegraphics[width=\textwidth]
    {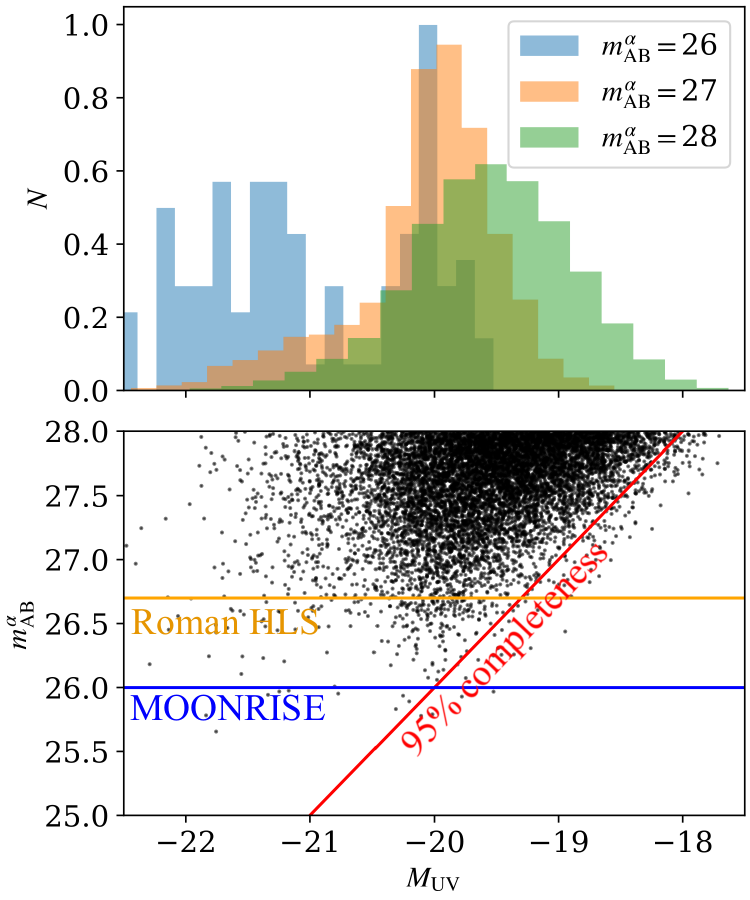}}
    \caption{Distribution of UV continuum magnitudes ($M_{\textrm{UV}}$) and 
    Lyman-$\alpha$ band AB magnitudes ($m_{\textrm{AB}}^\alpha$) of galaxies from 
    a $1$ Gpc on a side coeval cube at $z=7.2$. 
    The bottom panel includes only $10\%$ of the galaxies in the volume, for clarity. The horizontal lines indicate the limiting magnitudes for Roman HLS and the MOONRISE survey \citep{moonrise,hls2022}, and the diagonal line corresponds to
    the $M_{\rm UV}$ limit required for $95\%$ completeness at a given $m_{\rm AB}^\alpha$. The top panel shows the distributions of $M_{\rm UV}$ for 
    $m_{\rm AB}^\alpha=(26,27,28)$ in our model. We computed the $95\%$ completeness criterion 
    by integrating over these distributions.}
    \label{fig:mab_muv}
\end{figure}

For each pair of galaxy--21-cm instruments, we computed the S/N of the cross-power as a function of the galaxy survey area and AB magnitude limit in the 
relevant observing band, which is denoted as $m_{\rm AB}^\alpha$ for Ly-$\alpha$ selected 
surveys and $m_{\rm AB}$ for H$\alpha$/OIII selected surveys. Figure \ref{fig:mab_muv} illustrates the stochastic relation between intrinsic galaxy parameters and Ly-$\alpha$ 
emission, showing that the UV continuum magnitude of a galaxy $M_{\textrm{UV}}$ does not uniquely determine its $m_{\textrm{AB}}^\alpha$. We superimpose horizontal lines corresponding to the Ly-$\alpha$ band limiting AB magnitudes of the MOONRISE and Roman HLS surveys 
\citep{moonrise,hls2022}. Slit-spectroscopic galaxy surveys require a
photometric candidate survey, complete down to 
some $M_{\textrm{UV}}$, on which they perform follow-up observations. We did not investigate the effect of $M_{\textrm{UV}}$ incompleteness
of the photometric candidates on the assumption that a candidate survey is deep enough for a slit-spectroscopic follow-up survey of a 
given $m_{\textrm{AB}}^\alpha$ limit to be performed with full completeness. For example, a Lyman-$\alpha$ spectroscopic survey at $z\sim7$ with a depth of $m_{\textrm{AB}}^\alpha \leq 26$ would require follow-up of photometrically selected candidates down to $M_{\rm UV} \leq -20$ in order to be $\geq95\%$ complete (see Fig. \ref{fig:mab_muv}).

We produced six realizations of each observed galaxy field for 
magnitude cuts ranging from $m^\alpha_{\textrm{AB}}=25$ to 
$m^\alpha_{\textrm{AB}}=30$ and averaged the S/N at each magnitude cut. 
In our calculation of band magnitude, we assumed that the relevant band 
is dominated by line emission, which is reasonable because only 
LAEs with equivalent width greater than unity are detectable 
at the considered magnitude cuts. The total S/N of the 
cross-power spectrum varies fairly smoothly between magnitude cuts, 
so we devised an analytic form for S/N as a function of magnitude 
cut for each galaxy selection criterion, described 
in Appendix \ref{apdx:fov}. Moreover, the S/N varied trivially with 
the survey angular area and required only a cut in cylindrical Fourier 
space as well as the noise rescaling introduced in Equation \ref{eq:modes}. 
The redshift uncertainty is likewise easy to implement via Equation \ref{eq:pg}.

\begin{table*}[h]
    \centering
    \caption[]{\label{instruments}Benchmark galaxy surveys.}
       \begin{tabular}{lcccc}
          \hline \hline
          \textbf{Survey} & \textbf{Depth} ($m_{\textrm{AB}}$) & \textbf{Footprint} & $\bm{\delta}\bm{z}$ & \textbf{Redshift} \\ \hline
          SILVERRUSH (Subaru/Hyper Suprime-Cam) & $26$ & $16$ sq. deg. & $5\cdot10^{-2}$ & $z\sim6.6$ \\
          HLS (Nancy Grace Roman Space Telescope) & $26.7$ & $500$ sq. deg. & $10^{-2}$ & $z>7$ \\
          FRESCO (James Webb Space Telescope NIRCam grism) & $28.2$ & $62$ sq.arcmin. & $10^{-2}$ & $z>6$ \\
          MOONRISE (VLT MOONS) & $26$ & $1.2$ sq. deg. & $10^{-3}$ & $z>5$ \\
          a prospective ELT MOSAIC spectroscopic survey & $26-28$ & 1--3 sq. deg. & $10^{-3}$ & $z>5$ \\
          \hline
          \hline
       \end{tabular}
       \label{tab:surveys}
 \end{table*}

Although our predictions for the cross-spectrum S/N can 
be used to design new surveys, we used existing/proposed surveys 
as a reference. These are listed in Table \ref{tab:surveys}.
Of these, SILVERRUSH is already underway, and FRESCO, 
MOONRISE, and HLS are scheduled for execution in the coming years. 
We also consider a prospective spectroscopic 
follow-up survey using ELT MOSAIC modeled after MOONRISE.

The Systematic Identification of LAEs for Visible Exploration and 
Reionization Research Using Subaru HSC (SILVERRUSH) survey is 
a long-running program for the identification of tens of thousands 
of LAEs via narrow-band dropout covering redshifts $2-7$ 
\citep{silverrush2018}. The survey area of SILVERRUSH  ranges from 
$13.8$ square degrees at $z\sim5.7$ to $21.2$ square degrees at $z\sim6.6$ \citep{umeda24}. 
To split the difference between these survey footprints while producing 
the same number of galaxies, we took our SILVERRUSH benchmark to cover an equal $16$ 
square degrees at all redshifts and to observe down 
to $m^\alpha_{\textrm{AB}}=26$. The limiting factor when computing the S/N of the cross-power using SILVERRUSH is its narrow-band dropout strategy, which results in a relatively large redshift uncertainty. 
Many of these LAE are being followed up with spectroscopy, however, 
which should reduce these errors in the future.

The Nancy Grace Roman Space Telescope will conduct a High Latitude 
Survey (HLS) over a huge swathe of sky covering $\sim2000$ sq. deg. using its 
near-IR grism. Of this, $500$ sq. deg. overlap with the HERA stripe, and to facilitate the comparison between cross-correlation forecasts with SKA-low AA* and HERA-350, we limited the effective survey footprint of the Roman HLS to $500$ sq. deg.. Based on $\sim0.6$ years of observing time, the survey is expected 
to detect tens of thousands of $z>5$ galaxies via the Lyman-break technique. 
The frequency range of the near-IR band means that Roman can only detect Lyman-$\alpha$ 
emission at $z>7.2$ \citep{hls2022}.  \cite{la_plante23} forecast the significance of a cross-power spectrum detection made using a 
Lyman-$\alpha$ survey from Roman and a 21-cm field measured using HERA.  We confirm their findings and extend them to an SKA-low AA* measurement of the 21-cm signal.
Although HLS benefits from a large sky area, its imaging spectroscopy is 
limited to  $z>7.2$, which significantly limits the S/N compared to surveys extending to lower redshifts.  The LBG sample from Roman might 
be followed-up with slit spectroscopy (with, e.g. MOSAIC on the ELT), however, which would provide galaxy maps throughout the EoR.

The First Reionization Epoch Spectroscopically Complete
Observations (FRESCO) survey is a Cycle 1 medium program slated for execution 
on the James Webb Space Telescope \citep{fresco}. FRESCO will cover $62$ square 
arcminutes and observe out to a depth of $m_{\textrm{AB}}\approx28.2$ using NIRCam/grism. 
FRESCO targets H$\alpha$ at $z<7$ and [OIII] and H$\beta$ for $z\in(7,9)$. Although its 
target area is much smaller than that of the other benchmark surveys, FRESCO observes to much 
deeper magnitudes and selects on low-opacity emission lines. This allows a more accurate redshift determination (for simplicity, we used the same redshift uncertainty for all slitless surveys).

The MOONS Redshift-Intensive Survey Experiment (MOONRISE)
is a guaranteed observing time program of the MOONS instrument, a 
powerful spectrometer scheduled for deployment at the Very 
Large Telescope (VLT) \citep{moonrise}. MOONRISE will 
spectroscopically follow up $\sim1100$ candidate LAEs 
down to $m^\alpha_{\textrm{AB}}=26$ in a $\sim1$ square degree field at $z>5$. MOONRISE walks the middle 
path of reasonable survey area, moderate survey depth, and good redshift uncertainties, making 
it a promising candidate among slit spectroscopy surveys for cross-correlation with 21-cm. 

Finally, we provide forecasts using a prospective survey on the 
next-generation multiplex spectrograph ELT MOSAIC, which is currently 
in development for deployment at the Extremely Large Telescope
(ELT). We provide forecasts for MOONRISE-like surveys performed
using ELT MOSAIC assuming an $m^\alpha_{\textrm{AB}}=25.3$ $5\sigma$ 
5-hour limiting magnitude (\cite{mosaic}; C. Kehrig, private communication). 

\begin{figure}
    \resizebox{\hsize}{!}
       {\includegraphics[width=0.45 \textwidth]{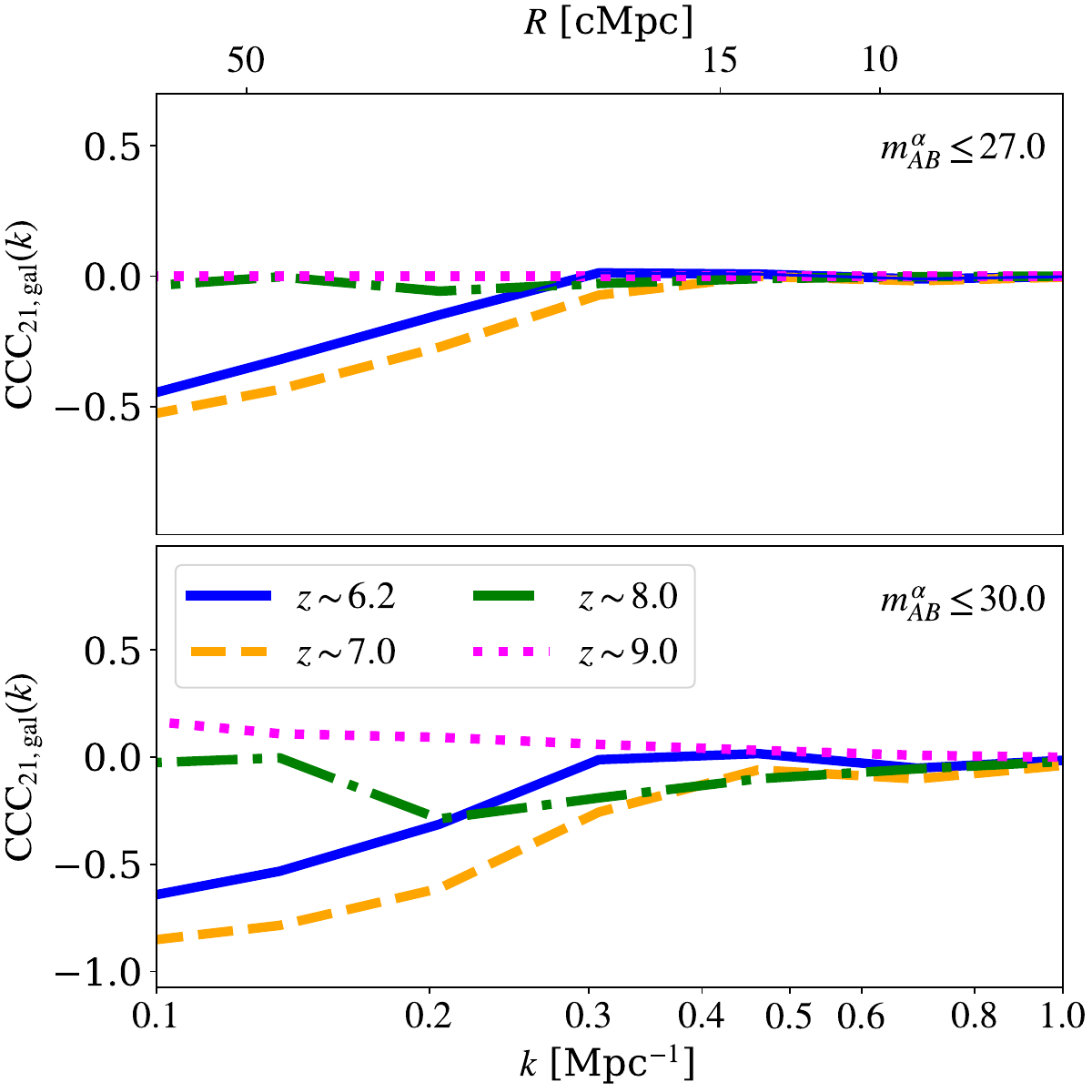}}
       \caption{
       Galaxy--21-cm correlation coefficients at $z\in(6.2,7.0,8.0,9.0)$, assuming at two Ly-$\alpha$ magnitude cuts: $m_{\rm AB}^\alpha\leq27$ ({\it top panel}) and $m_{\rm AB}^\alpha\leq30$ ({\it bottom panel}).
       Our fiducial model at $z=6.2$, $7.0$, $8.0$, and $9.0$ results in a mean IGM neutral fractions of
       $\bar{x}_{\rm HI}=0.15$, $0.41$, $0.61$, and $0.79$, 
       and mean HI spin temperature-to-CMB temperature 
       ratios of $\bar{T}_{\rm S}/T_{\gamma}=10.4$, $ 4.72$, $2.10$, and $0.79$, respectively.
       On the top horizontal axis we show the comoving scale corresponding to a given wavemode:  $R=2\pi/k$.}
       \label{fig:ps1d}
\end{figure}

\begin{figure*}
    \resizebox{\hsize}{!}
       {\includegraphics[width=0.45 \textwidth]{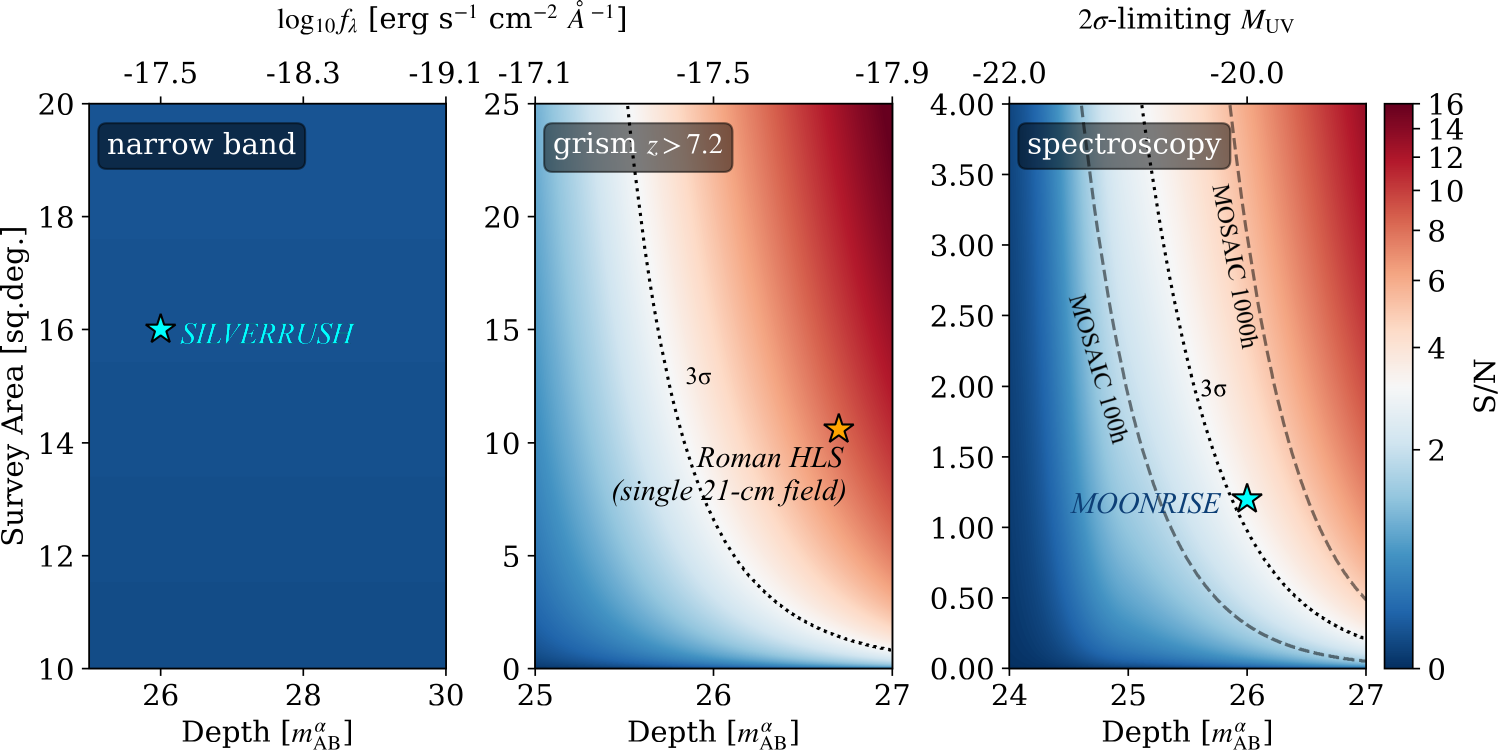}}
       \caption{Effect of observing depth, field of view, and galaxy survey
       type on the signal-to-noise ratio of a cross-power spectrum detection.
       Each panel corresponds to a different galaxy survey type, which determines the mean redshift uncertainty associated with that survey. On the top axis of the slit spectroscopy panel, we show the corresponding rest frame UV magnitude limit required for $\geq 2\sigma$ completeness of the photometric candidate sample  (see Fig. \ref{fig:mab_muv}). Blue stars correspond to the SILVERRUSH and MOONRISE high-redshift galaxy surveys, and the orange star corresponds to an overlap of the Roman HLS survey and a single SKA-low AA* beam; multiple 21-cm fields can be tiled to increase the S/N in quadrature as discussed in the text. The MOONRISE spectroscopic survey should enable a $3\sigma$ detection of the cross-spectrum. In the spectroscopy panel we also show  $100$- and $1000$-hour isochrones for potential future surveys using MOSAIC on the ELT.}
       \label{fig:depth-scaling}
\end{figure*}

\begin{figure*}
    \resizebox{\hsize}{!}
       {\includegraphics[width=0.45 \textwidth]{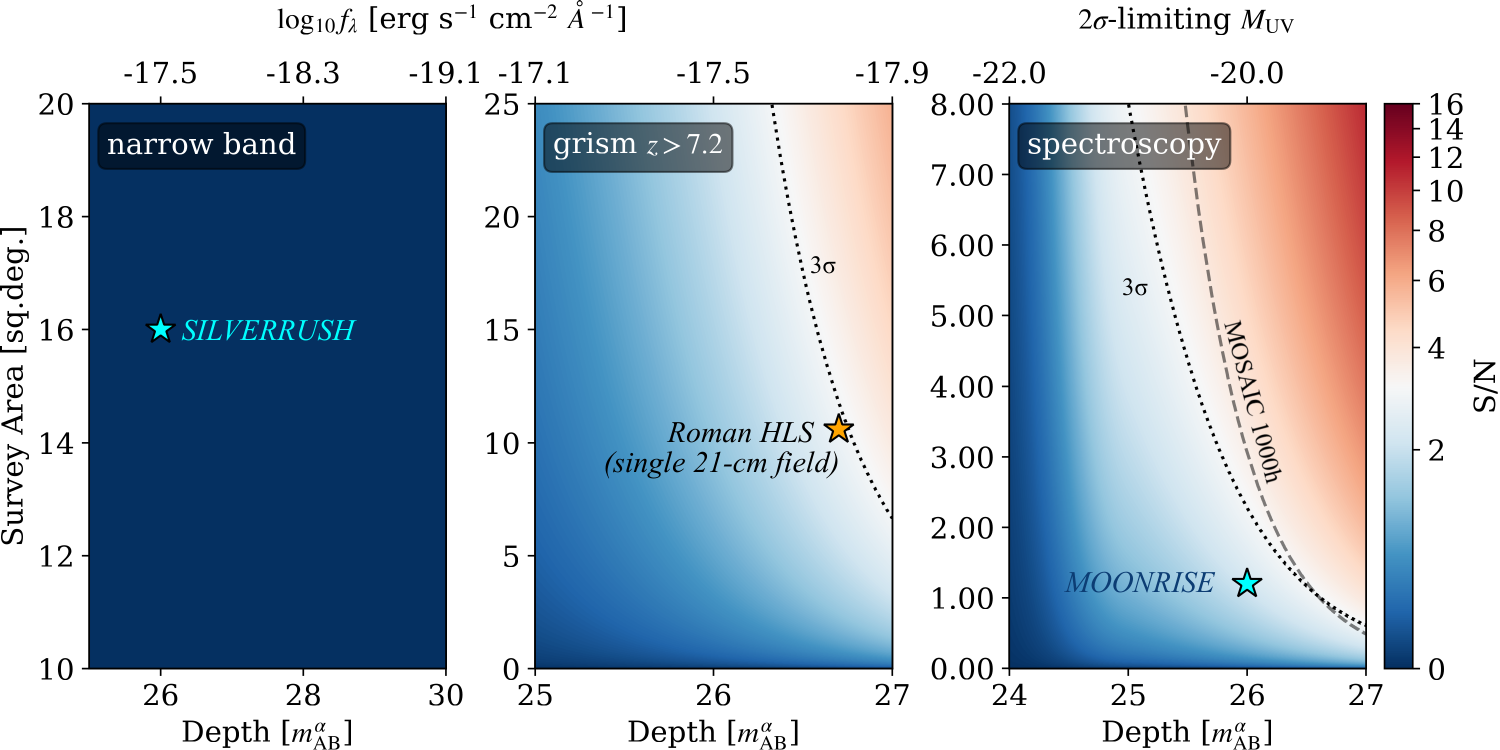}}
       \caption{As Figure \ref{fig:depth-scaling}, but assuming the pessimistic 
       foregrounds shown in Figure \ref{fig:megaplot}. Here, MOONRISE is no longer sufficient for a $3\sigma$ cross-spectrum detection, but a $1000$-hour survey with MOSAIC remains viable.}
       \label{fig:depth-scaling-pessimistic}
\end{figure*}

\begin{figure*}
    \resizebox{\hsize}{!}
       {\includegraphics[width=0.45 \textwidth]{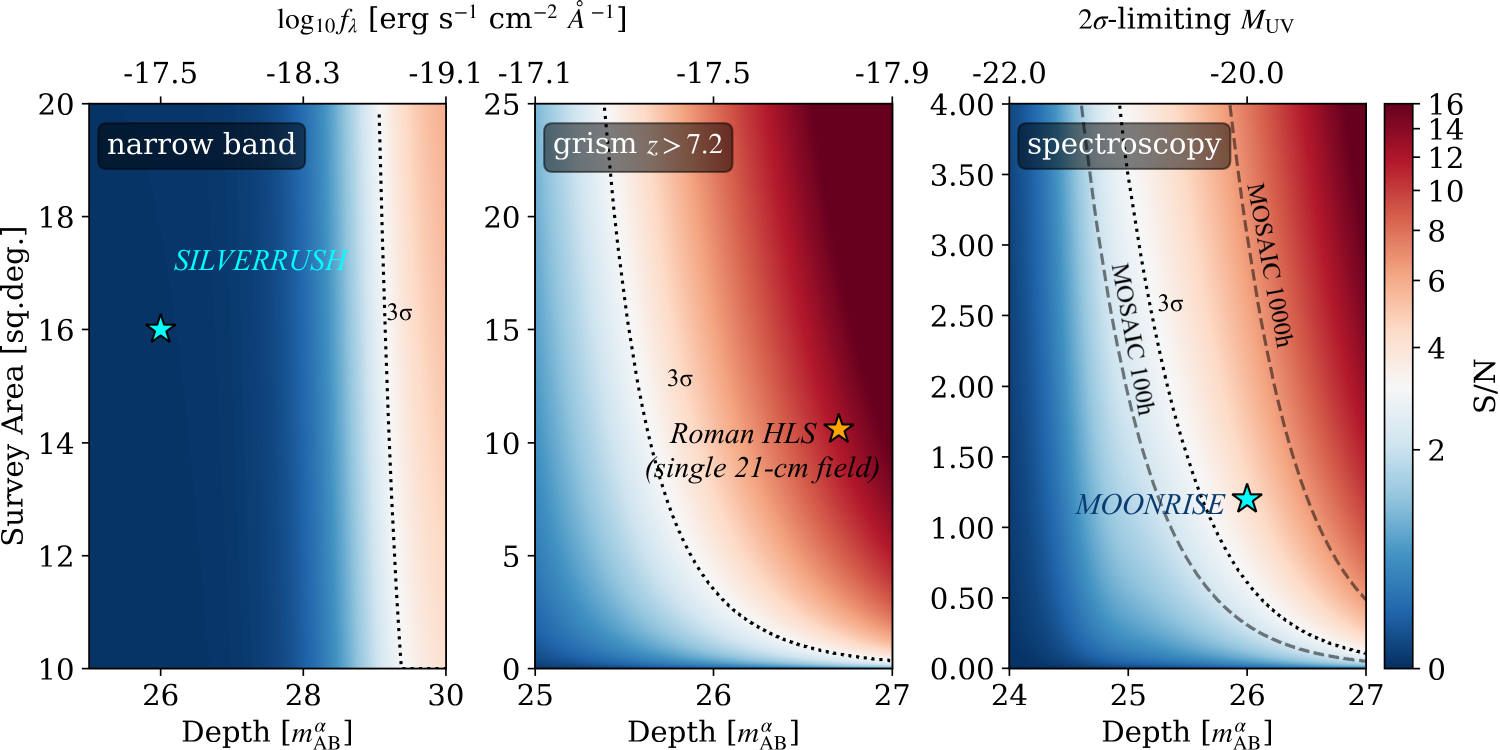}}
       \caption{As Figure \ref{fig:depth-scaling}, but assuming the optimistic foreground scenario in Figure \ref{fig:megaplot}. 
       SILVERRUSH remains not viable, although an extremely deep survey of similar angular extent could theoretically breach 3$\sigma$, 
       and the S/N with spectroscopy improves only slightly.}
       \label{fig:depth-scaling-recovery}
\end{figure*}

\begin{figure*}
    \resizebox{\hsize}{!}
       {\includegraphics[width=0.45 \textwidth]{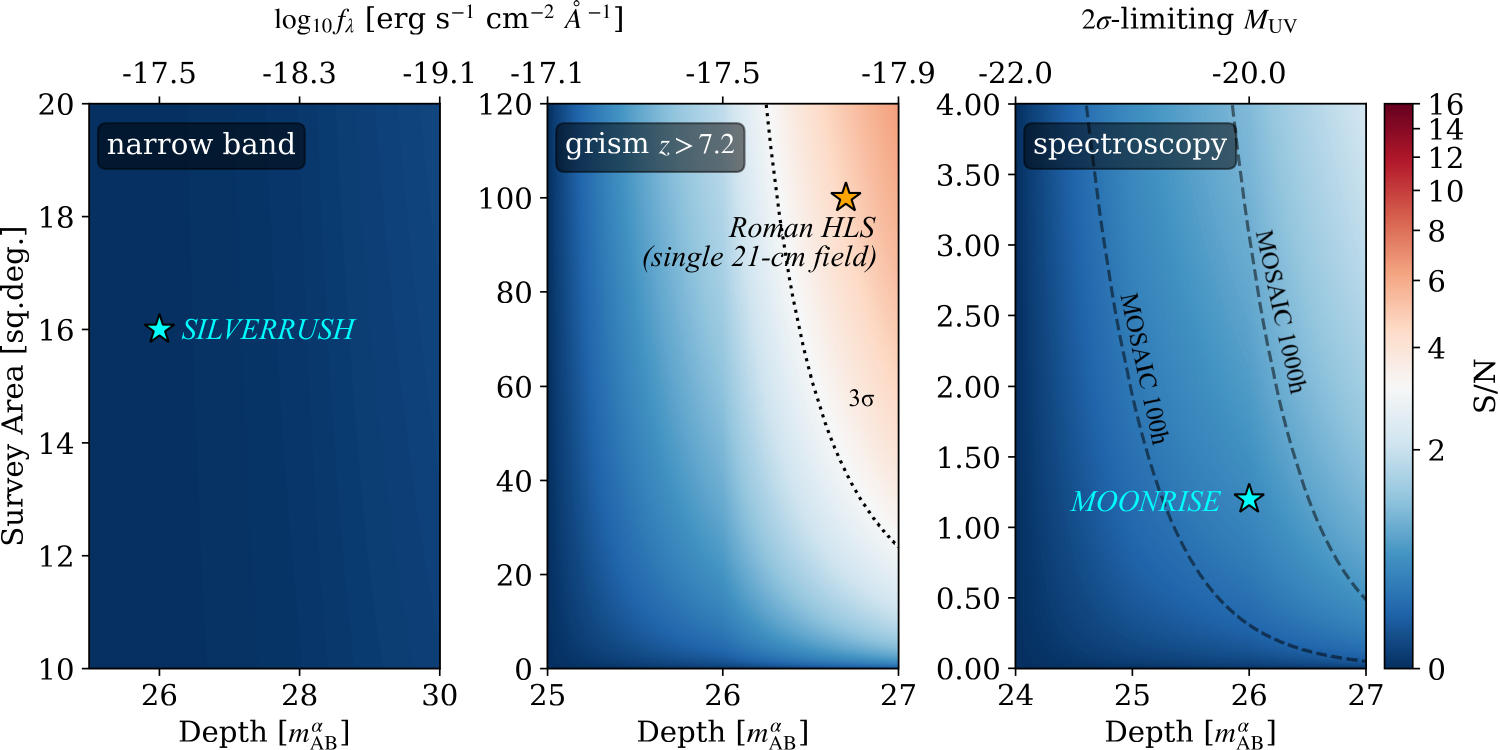}}
       \caption{As Figure \ref{fig:depth-scaling}, but using HERA rather than the SKA-low as the fiducial 21-cm interferometer. 
       Notably, the lack of sensitivity of HERA to high $k_\perp$ modes (i.e. small sky plane scales) precludes detection of the cross-spectrum using a galaxy field with a small ($\sim5$ sq. deg.) field of view.  Wide-area slitless spectroscopy remains the only viable candidate for cross-correlating galaxy maps with HERA 21-cm data.}
       \label{fig:depth-scaling-hera}
\end{figure*}

\begin{figure}
    \resizebox{\hsize}{!}
       {\includegraphics[width=0.45 \textwidth]{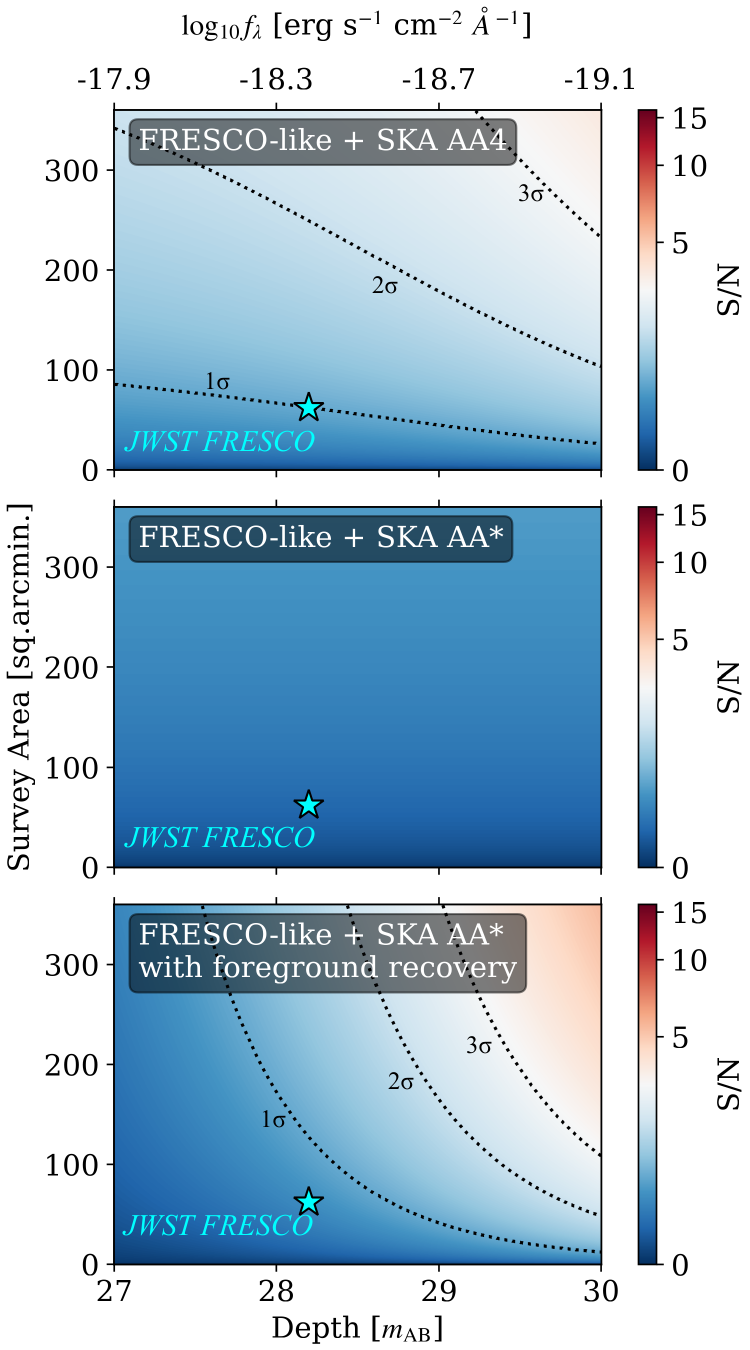}}
       \caption{Effect of observing depth and field of view on a FRESCO-like survey that selects for H$\alpha$ at $z<7$ and [OIII]/H$\beta$ at $z\geq7$. The FRESCO $62$ square arcminute fields are too small to yield a detection of the 21-cm cross-power spectrum.}
       \label{fig:depth-scaling-fresco}
\end{figure}

\section{Results}
\label{sec:disc}

Before presenting our main results of the S/N forecasts, 
we briefly review the general behavior of the cross-power 
spectrum and the information it contains. Figure 
\ref{fig:ps1d} shows the galaxy--21-cm cross correlation 
coefficients (CCCs) at $z\in(6.2,7.0,8.0,9.0)$ for our fiducial model.  The CCCs are defined as

\begin{equation}
    \rm{CCC}_{21,g}(k)=\frac{P_{21,g}(k)}{\sqrt{P_{21}(k)P_{g}(k)}},
\end{equation}

\noindent
where $P_{21,g}(k)$, $P_{21}(k)$, and 
$P_{g}(k)$ are the 1D galaxy--21-cm cross-, 
21-cm auto-, and galaxy auto-power spectra, 
respectively. With this normalization, the CCCs provide a clean measure of the phase difference between the two fields: CCC $\rightarrow0$ for uncorrelated fields, CCC $\rightarrow1$ for correlated fields, and CCC $\rightarrow-1$ for anti-correlated fields 
(e.g., \citealt{lidz09, sobacchi16, fronenberg24}).
The top and bottom panels correspond to different Ly-$\alpha$ 
magnitude cuts. One cut is ambitiously deep ($m_{\rm AB}^\alpha\leq27$; cf. Table \ref{tab:surveys}) and another cut is even deeper in order to show physical trends ($m_{\rm AB}^\alpha\leq30$). 

We recovered the qualitative trends noted in previous studies on the cross-power and the CCCs (e.g., \citealt{furlanetto07, lidz09, sobacchi16, hutter23, la_plante23}).  The CCCs tend toward zero on small scales during the EoR ($z\lesssim 8$) and the preceding EoH ($z\gtrsim 8$; see Figure  \ref{fig:cutout}) because the relative locations of galaxies inside either ionized HII bubbles or heated IGM patches do not impact the cross-power, and thus the two fields on those small scales could have random phases (e.g., \citealt{lidz09, sobacchi16, fronenberg24}).  On the other hand, because the galaxies and 21-cm both depend on the large-scale matter field, the CCCs on large scales tend to either -1 or 1, during the EoR and EoH, respectively (e.g., \citealt{heneka20}).  The transition of the CCCs from 0 to 1 (-1) encodes the typical sizes of heated (ionized) IGM patches during the EoH (EoR; see \citealt{sobacchi16, heneka20, moriwaki24}).

However, the asymptotic trend of the CCC to 1 or -1 on large scales is only reached in the limit of an infinite galaxy survey depth, which would effectively measure the DM halo--21cm cross-correlation (see Fig. 3 in \citealt{sobacchi16}). Realistic magnitude cuts together with the intrinsic stochasticity in the galaxy -- halo connection discussed in the previous section, result in a much smaller large-scale (anti)correlation of the galaxy and 21cm fields.   For $m_{\rm AB}^\alpha\leq27$ in the top panel of Fig. \ref{fig:ps1d}, the CCCs only reach -0.5 during the EoR, while no signal is detected during the EoH at $z>8$ (see also Fig. \ref{fig:cumul_snr}).  For $m_{\rm AB}^\alpha\leq30$ in the bottom panel, the large-scale CCC trends are closer to the theoretically expected limiting cases, and even the positive correlation during the EoH can be seen at $z=9$.  However, depths of $m_{\rm AB}^\alpha\leq30$ on reasonably sized fields would require over 10k hours of integration with slit spectroscopy, as we describe below.

In Figures \ref{fig:depth-scaling}-
\ref{fig:depth-scaling-fresco} we show the main results of this work: The expected S/N 
for every survey configuration considered in this study. Each figure shows the S/N of the galaxy--21-cm cross-power spectrum as a function of observing depth and survey footprint, for narrow-band dropout, slitless and slit spectroscopy ({\it left to right panels}).  Different figures vary the 21-cm instrument (HERA-350 vs SKA-low AA*) as well as the level of 21-cm wedge contamination (optimistic, fiducial, pessimistic).  In the spectroscopy panels for Ly$\alpha$-selected 
galaxy surveys, we show $100$- and $1000$-hour isochrones 
for observations using the ELT MOSAIC instrument. 
These isochrones correspond to the total observation time, equal to the number of 40 sq. arcmin pointings required to reach the desired survey footprint on the vertical axis, times the exposure time per pointing required for a 
$\geq 5\sigma$ detection down to the given limiting $m^\alpha_{\textrm{AB}}$ on the horizontal axis (using the latest MOS-NIR-LR sensitivity estimates; C. Kehrig, private communication).

An increased redshift uncertainty 
suppresses S/N exponentially, which requires an 
exponential rise in the number of galaxies that are 
detected to compensate. This may be achieved by either 
dramatically increasing the survey footprint or increasing the 
observing depth by a few magnitudes.
Therefore, we evaluated the S/N of narrow-band dropout and grism 
surveys on a wider range of footprint sizes and a deeper 
range of observing depths than the slit spectroscopic 
surveys. Furthermore, we limited the sensitivity 
of the grism surveys in this experiment to $z>7.2$
to reflect the grism filter of the Nancy Grace Roman 
Space Telescope.

When the surveys are deeper than $m^\alpha_{\rm AB} \sim 25$, 
the S/N increases more readily with survey area than with the observing depth in all cases. 
For spectroscopic surveys, we find that targeting a depth of $m^\alpha_{\rm AB}\sim26$ 
and maximizing the contiguous survey footprint yields the highest S/N per unit survey time. 

In our fiducial model, $m^{\alpha}_{\rm AB}\sim26$ 
corresponds to the observing depth at which the 
maximum value of the galaxy power spectrum 
$P_{\rm gal}(k)$ roughly equals the noise power 
$1/n_{\rm gal}$ (ignoring redshift uncertainty), 
and therefore, the S/N of the galaxy auto-power spectrum 
at this depth for a single field is $\sim1$. This is 
consistent with the recommendation of \cite{tegmark97} 
to extend the depth of a single field until the S/N reaches 
$1$ and then tile additional fields to 
time-efficiently maximize the total S/N.
We discuss this result in greater detail in Appendix \ref{apdx:fov}.

\subsection{SKA--Lyman Alpha}

Figure \ref{fig:depth-scaling} shows our results for 
a Lyman-$\alpha$ galaxy survey paired with an SKA-low AA*
21-cm measurement assuming the fiducial foreground scenario. 
The left panel shows that the SILVERRUSH sample without a 
spectroscopic follow-up is insufficient for a cross-power detection.  
This contradicts some previous claims (e.g., \citealt{sobacchi16, hutter18b}), which treated the redshift uncertainties in the 
LAE sample more simplistically. Even an extremely deep 
narrow-band dropout survey (i.e., $m^\alpha_{\textrm{AB}}<30$) of 
a similar area cannot yield a significant cross-spectrum detection, 
except in the case of optimistic 21-cm foreground recovery. This 
highlights the importance of small redshift uncertainties 
in measuring the cross-power spectrum. 

The middle panel includes a forecast for Roman HLS using a 
survey footprint equivalent to the beam size of SKA-low AA*. The 
enormous angular survey extent of Roman means that in this case 
alone, the footprint for the cross-correlation is limited by the 
field of view of the 21-cm instrument. We find that a single patch of 
the SKA-low AA* beam within Roman HLS yields an $8\sigma$ detection of 
the cross-power spectrum. Tiling multiple beams can increase the S/N. 
For example, $47$ patches of the SKA-low AA* beam can fit within 
$500$ sq. deg. of overlapping area, which would bring the cumulative S/N from to $\sqrt{47}\cdot8=54.8$.

The right panel shows that the planned MOONRISE survey should be 
sufficient to measure the cross-power spectrum at a $\sim3\sigma$ significance. 
Furthermore, the dashed contours delineating prospective MOSAIC configurations 
imply that a similar survey using MOSAIC could achieve a $\sim4\sigma$ measurement 
of the cross-power spectrum in $\sim500$ hours of observation targeting $\sim$1--3 sq. deg. An extension of this to $1000$h can bring the S/N up to $5\sigma$.

Figures \ref{fig:depth-scaling-pessimistic} and 
\ref{fig:depth-scaling-recovery} show our results for 
the same Lyman-$\alpha$ galaxy survey and SKA-low observations,
but assuming pessimistic and optimistic 21-cm foregrounds, 
respectively. Figure \ref{fig:depth-scaling-pessimistic} shows 
that the significance of a cross-spectrum 
measurement depends very strongly on the level of the 
foregrounds and that our strong foreground case 
precludes a detection of the cross-power spectrum using the 
MOONRISE survey. However, a $1000$-hour observation using 
ELT MOSAIC could yield a $\sim3\sigma$ detection. We find that 
an increase in the number of telescope 
pointings (assuming equal exposure time per pointing) yields more S/N per unit 
observing time than boosting exposure depth. 
Figure \ref{fig:depth-scaling-recovery} shows that when 
a significant portion of 21-cm foreground wedge modes 
is recovered, the narrow band dropout redshift uncertainty 
still precludes a cross-spectrum measurement for the range of survey configurations we considered.

\subsection{HERA--Lyman-alpha}

Figure \ref{fig:depth-scaling-hera} shows our results for a 
Lyman-$\alpha$ galaxy survey paired with a measurement of 
the 21-cm signal made using HERA-350. 
Compared with SKA-low, the lack of sensitivity of HERA to high $k_\perp$ modes precludes cross-spectrum detections using galaxy fields 
with small angular footprints.  This is evident in the rightmost panel of Figure \ref{fig:depth-scaling-hera}, where 
neither MOONRISE nor MOSAIC can yield a cross-spectrum detection with HERA.

The most promising candidate for a cross-correlation with HERA are 
large-area surveys with slitless spectroscopy, shown 
in the middle panel using the benchmark HLS with Roman grism. 
The star in this panel shows the forecast from \cite{la_plante23}, 
which we match to within $10\%$\footnote{
It is curious, given our different 
methods, that we match the predicted S/N 
of a HERA-Roman survey so closely. Relative 
to the model of \cite{la_plante23}, we produce a lower 
number density of highly luminous Ly-$\alpha$-emitting 
galaxies, and we therefore expect a lower 
S/N in the galaxy power spectrum for equal observing 
depth. However, our 21-cm noise model, 
\texttt{21cmSense}, differs from the analytic 
model of \cite{la_plante23} in that it accounts for
baseline rotation during the observation, lowering the 
noise level in certain $k$-bins relative to the 
analytic model. In both cases, the deviations between 
our model and that of \cite{la_plante23} are $\sim20\%$, 
and they conspire to offset each other such that 
our estimated S/N matches their model within $\sim10\%$.
}.
We predict a $\gtrsim6 \sigma$ detection of the cross-power for a 
Roman HLS + HERA-350 observation in a single HERA beam covering 
$100$ sq. deg. Since HERA always operates in drift-scan mode, 
it will observe a cumulative five patches of this size in the 
Roman HLS field, which means a total of S/N$=\sqrt{5}\cdot6\approx13.4$ 
for the full $500$ sq. deg. of overlap. This is very close to the 
S/N=$14$ quoted by \cite{la_plante23}.

\subsection{SKA--OIII/H Alpha}

Figure \ref{fig:depth-scaling-fresco} shows our 
results for a NIRCam-like galaxy survey targeting H$\alpha$ 
below $z=7$ and [OIII]/H$\beta$ for $z\in(7,9)$ paired 
with a 21-cm signal measured by the SKA-low. The two bottom panels 
show S/N forecasts for a cross-correlation using SKA-low AA*, 
and the top panel uses SKA-low in its planned AA4 upgraded
configuration, which includes additional outrigger antennas 
\citep{skao23}. The two top panels assume fiducial foregrounds, 
and the bottom panel uses optimistic foregrounds. Fiducial 
foregrounds preclude a detection of the 
galaxy--21-cm cross-power spectrum in a FRESCO-like survey paired with 
SKA-low AA*. Even when we invoke AA4 or optimistic foregrounds, 
the FRESCO footprint remains too small to claim a cross-spectrum 
detection. However, a FRESCO-like survey with the same observing depth 
but ten times the survey footprint would yield a $\sim3\sigma$ detection 
of the 21-cm signal in cross-correlation using SKA-low AA4. This is not 
far-fetched because FRESCO itself is only a $53$-hour program and 
produces two noncontiguous pencil beams. Using the same exposure time
per pointing, it would only take $\sim240$ hours to expand a single FRESCO beam into a field that is large enough for cross-correlation.

\subsection{Redshifts that contribute to the signal-to-noise ratio}

\begin{figure}
    \resizebox{\hsize}{!}
       {\includegraphics[width=0.45 \textwidth]{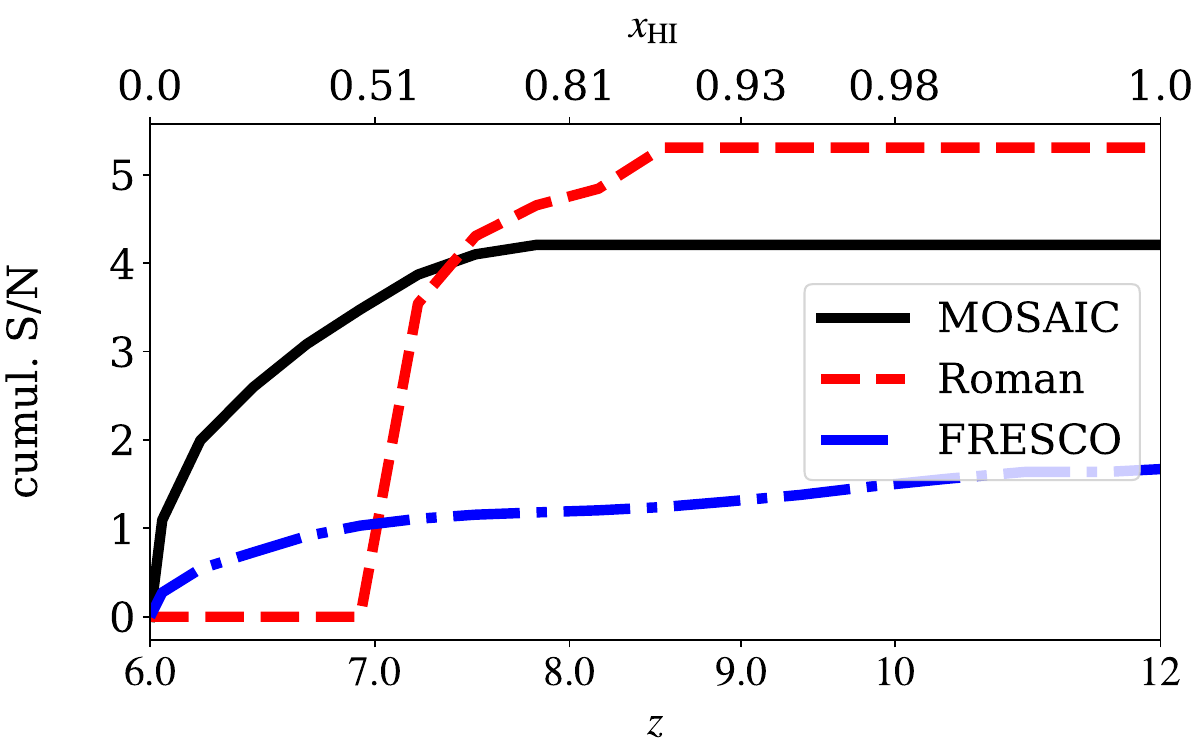}}
       \caption{
       Cumulative S/N of the galaxy--21-cm cross-power spectrum as a function of redshift for fiducial observations using MOSAIC, JWST, and Roman. The MOSAIC observation corresponds to our recommended ELT survey, the Roman observation corresponds to and integration over two patches of the SKA-low field of view within the Roman observing area, and the JWST observation corresponds to the FRESCO survey. The Roman and MOSAIC surveys are Ly-$\alpha$ selected, so that redshifts higher than $\sim8$ contribute negligibly to their total S/N due to attenuation from the neutral IGM.  On the other hand, FRESCO selects on H$\alpha$ and [OIII], allowing higher-redshift galaxies to contribute to the total S/N. The global neutral IGM fraction is indicated on the upper horizontal axis.}
       \label{fig:cumul_snr}
\end{figure}

We also investigated which redshift bins contribute 
the most to the total S/N. We show the cumulative S/N 
for fiducial galaxy surveys using MOSAIC, Roman, and
JWST (with FRESCO-like selection) as a function of
redshift (bottom axis) and IGM neutral fraction 
(top axis) in Figure \ref{fig:cumul_snr}. 
The lowest-redshift bins with 
nonzero neutral hydrogen dominate the S/N of all 
surveys we considered. This reflects the 
fact that any magnitude-limited survey detects significantly more 
galaxies at low redshift than at high redshift. This 
effect is most pronounced for Ly-$\alpha$ selected 
surveys because the sensitivity of Ly$\alpha$ to IGM attenuation 
increases the cross-power during the EoR while making it more difficult 
to detect galaxies pre-EoR. Galaxy surveys targeting nebular lines with 
low IGM opacities, such as FRESCO, result in an S/N that is distributed 
over a broader range in redshift.  The plateau in the cumulative S/N at 
$z\sim8$--9 is caused by the equality between the 21-cm spin temperature 
and the CMB temperature at this redshift in 
our fiducial model, which drives a sign change in the cross-power 
(see Figure \ref{fig:ps1d} and associated discussion).

\section{Conclusion}
\label{sec:conc}

A detection of the cosmic 21-cm signal would usher in a new era for 
research on the Epoch of Reionization (EoR). 
Radio interferometers aim to make a preliminary detection of the power 
spectrum with a low signal-to-noise (S/N) in 
the coming years. A cross-correlation of the preliminary
signal with galaxies will provide a crucial sanity 
check for initial 21-cm detection claims 
because the tracer systematics are uncorrelated.

We forecast the S/N of the cross-power in various pairings of 
prospective galaxy surveys and 21-cm interferometers. Our fiducial 
21-cm observations used $1000$-hour integrations of 
the SKA-low AA* and HERA-350 instruments. These were paired with galaxy surveys of 
varying survey area, depth, and selection method 
(narrow-band, slitless spectroscopy, and slit spectroscopy) to 
compute the galaxy--21-cm cross-power spectrum and quote an 
S/N for each scenario. We also varied the level of 21-cm foreground 
contamination that would be excised before computing the cross-power.

To compute mock observations, we self-consistently simulated 
large ($1$ Gpc) galaxy and 21-cm fields using \texttt{21cmFASTv4} 
\citep{davies25}. We then post-processed the galaxy field to include 
only galaxies whose line luminosities in a given band (spanning either 
Lyman-$\alpha$ or H$\alpha$/H$\beta$/OIII) exceeded the sensitivity 
threshold for the survey of interest. To connect the line luminosities to the galaxy UV continuum magnitude or star formation rate, we sampled empirical relations provided by \cite{mason2018} for Ly-$\alpha$ and \cite{lim17} for other lines.

Our mock galaxy surveys were set up for an easy comparison with 
benchmark instruments, including Subaru HyperSupremeCam, Roman grism, 
VLT MOONS, ELT MOSAIC, and JWST NIRCam. Our principal results are listed 
below.

\begin{itemize}
    \item Even pessimistic 21-cm foregrounds do not preclude a detection of the cross-power with galaxies when the  slitless spectroscopic survey area is very large or the slit spectroscopic survey is very deep.  Overall, the different foreground contamination scenarios we studied affect the achievable S/N by factors of few.
    \item Narrow-band dropout surveys are unlikely to detect the cross-power spectrum due to their poor redshift localization.
    \item Large-field slitless spectroscopy with Roman grism yields the highest 
    S/N cross-power detection: $\sim55\sigma$ ($\sim13\sigma$) paired with SKA-low AA* (HERA-350), for our fiducial model and assuming $\sim$500 sq. deg. of overlap.  Although the Roman grism is only sensitive to Ly-$\alpha$ at $z\geq7.2$, the Roman 
     redshift precision and huge observing area more than compensate for the loss 
    in signal from the missing lower-redshift galaxies.
    \item Small-field slitless spectroscopy targeting H$\alpha$/H$\beta$OIII using JWST NIRCam, as per the FRESCO survey \citep{fresco}, does not yield 
    a high enough S/N for a detection of the cross-power spectrum.  However, a 
    survey of similar depth that covers about ten times the area would yield a $\sim3\sigma$ detection of 
    the 21-cm signal in cross-correlation with a 21-cm measurement provided by a planned upgrade of SKA: AA4.
    \item Slit spectroscopy can provide high S/N cross-power for SKA-low AA* observations. Specifically, the planned MOONRISE survey with VLT MOONS can result in a $\sim3\sigma$ detection \citep{moonrise}. We likewise showed that a $4\sigma$ detection can be made with ELT MOSAIC for a comparable observation time for a survey with area $\sim$1--3 sq. deg. \citep{mosaic}.
\end{itemize}

\noindent These forecasts are intended as a guide for future survey strategies to facilitate the detection of the galaxy--21-cm cross-power spectrum.

\begin{acknowledgements}
    We thank S. Murray for helpful guidance regarding the generation of noise for the 21-cm field. 
    Additional thanks to D. Breitman for providing a handy script for rapid power spectrum calculations.
    We extend further gratitude to S. Carniani, L. Pentericci, and M. Hayes for 
    useful discussions regarding upcoming surveys, as well as to A. C. Liu for invaluable feedback on a draft version of this paper. 
    We gratefully acknowledge computational resources of the HPC center at SNS.
    AM acknowledges support from the Italian Ministry of Universities and Research (MUR) through the PRIN project "Optimal inference from radio images of the epoch of reionization", and the PNRR project "Centro Nazionale di Ricerca in High Performance Computing, Big Data e Quantum Computing".
\end{acknowledgements}

\bibliography{gxcor}{}
\bibliographystyle{aa}

\begin{appendix}

\onecolumn

    \section{IGM Optical Depth to Lyman-alpha}
    \label{apdx:tau}

    The above relations for computing Ly-$\alpha$ luminosity from $M_{UV}$ were calibrated 
    against observation from redshifts $3-4.6$, a time when the IGM was already ionized, 
    meaning that while they include the effects of attenuation from HI in the ISM/CGM, 
    they do not include any attenuation from the IGM \citep{oyarzun17}. To account for 
    the optical depth of the IGM at our redshifts of interest, we adapted the model of 
    \cite{mesinger2008}, which provided analytic relations for the IGM attenuation of 
    rest-wavelength Ly-$\alpha$ emission for halos of a given mass and mean ionization 
    fraction at $z=9$. We opted to use this 
    analytic relation for simplicity's sake, deferring a realistic treatment 
    of Ly-$\alpha$ attenuation along sightlines in the IGM to a future work. 
    \cite{mesinger2008} found that the distribution of IGM optical depths for a 
    halo of a given mass in a region of given mean neutral fraction is well-described by the log-normal distribution 
    
    \begin{equation}
        \frac{\partial p(>\tau_{\textrm{IGM},9})}{\partial\tau_{\textrm{IGM},9}}=\frac{1}{\tau_{\textrm{IGM},9}\sqrt{2\pi\sigma^2_{\textrm{IGM},9}}}\exp\left[-\frac{(\ln\tau_{\textrm{IGM},9}-\mu_{\textrm{IGM},9})^2}{2\sigma^2_{\textrm{IGM},9}}\right],
    \end{equation}
    
    \noindent where
    
    \begin{equation}
        \mu_{\textrm{IGM},9}=-3.37+\log_{10}M_{10}(-0.115-0.587\bar{x}_{\textrm{HI}})+5.30\bar{x}_{\textrm{HI}},
    \end{equation}
    
    \begin{equation}
        \sigma_{\textrm{IGM},9}=1.68+\log_{10}M_{10}(-0.115-0.285\bar{x}_{\textrm{HI}})-1.08\bar{x}_{\textrm{HI}},
    \end{equation}
    
    \noindent wherein $M_h=M_{10}\cdot10^{10}M_\odot$ is the halo mass and $\bar{x}_{\textrm{HI}}$ is the mean neutral fraction of the IGM surrounding the halo.
    
    We extend this model in two ways, firstly by generalizing to other redshifts, and secondly by accounting for the frequency-dependence of the IGM optical depth. We account for the first effect simply by scaling the optical depth with the mean density of the Universe, i.e.,
    
    \begin{equation}
        \tau_{\textrm{EoR}}(z)=\tau_{\textrm{IGM},9}\left(\frac{1+z}{10}\right)^3.
    \end{equation}
    
    \noindent For the latter effect, we use the Miralda-Escudé equation for the Ly-$\alpha$ damping wing and normalize it such that its value at rest wavelength equals the value provided by \cite{mesinger2008} \citep{miralda98}. The Miralda-Escudé equation is 
    
    \begin{equation}
        \tau_{\textrm{ME}}(z)=\tau_{\textrm{GP}}\frac{R_\alpha}{\pi}\left(\frac{1+z_b}{1+z}\right)^{1.5}\left[I\left(\frac{1+z_b}{1+z}\right)-I\left(\frac{1+z_e}{1+z}\right)\right],
    \end{equation}
    
    \noindent where $\tau_\textrm{GP}=7.16\cdot10^5((1+z)/10)^{1.5}$ is the 
    Gunn-Peterson optical depth \citep{gp65}, $R_\alpha=6.25\cdot10^8/(4\pi\nu_{\textrm{Ly}\alpha})$ 
    is the decay scale of Ly-$\alpha$ emission, wherein $\nu_{\textrm{Ly}\alpha}$ is the frequency of 
    Ly-$\alpha$ emission, $z=(\lambda/\lambda_{\textrm{Ly}\alpha})(1+z_s)-1$ is the redshift 
    corresponding to a rest wavelength $\lambda$, wherein $\lambda_{\textrm{Ly}\alpha}$ is the 
    rest wavelength of the Ly-$\alpha$ line, $z_s$ is the systemic redshift of the LAE, $z_b$ 
    is the redshift whereat the Ly-$\alpha$ emission first interacts with the neutral IGM, 
    taken in this study to be $z_b=z_s-0.02$, $z_e$ is the redshift whereat the Ly-$\alpha$ 
    emission exits the neutral IGM, taken to be $z_e=5.5$, and finally
    
    \begin{equation}
        I(x)\equiv\frac{x^{9/2}}{1-x}+\frac{9}{7}x^{7/2}+\frac{9}{5}x^{5/2}+3x^{3/2}+9x^{1/2}-\ln\left|\frac{1+x^{1/2}}{1-x^{1/2}}\right|
    \end{equation}
    
    \noindent accounts for the shape of the Ly-$\alpha$ transmission curve.
    
    Combining all parts, our equation for the Ly-$\alpha$ optical depth as a function of rest-frame wavelength and systemic redshift is
    
    \begin{equation}
        \tau_{\textrm{EoR}}(\lambda)=\tau_{\textrm{IGM},9}\frac{\tau_{\textrm{ME}}(\lambda)}{\tau_{\textrm{ME}}(z_s)}\left(\frac{1+z_s}{10}\right)^3.
    \end{equation}
    
    \section{Signal-to-Noise Ratio as a Function of Observing Depth}
    \label{apdx:fn}

    We produce each panel of Figures \ref{fig:depth-scaling}, \ref{fig:depth-scaling-pessimistic}, and \ref{fig:depth-scaling-recovery} using only three realizations of the biased galaxy field. For both galaxy selection criteria, we generate mock observations with $m_{\textrm{AB}}$ sensitivities of $25$, $26$, $27$, $28$, $29$, and $30$. In order to produce a S/N gradient across the full range of magnitudes in each panel, we require a means of interpolating the S/N between each realization. 
    
    We adopt a double sigmoid function of the form 
    
    \begin{equation*}
        \frac{\textrm{S}}{\textrm{N}}(m_{\textrm{AB}}) = 
        \begin{cases} 
              h_1(1+\exp\{-k_1(m_{\textrm{AB}}-r_1)\})^{-1} & m_{\textrm{AB}}\leq 25 \\
              h_2(1+\exp\{-k_2(m_{\textrm{AB}}-r_2)\})^{-1} & m_{\textrm{AB}}> 25
       \end{cases}.
    \end{equation*}

\begin{figure}
        \centering
        {\includegraphics[width=\textwidth]{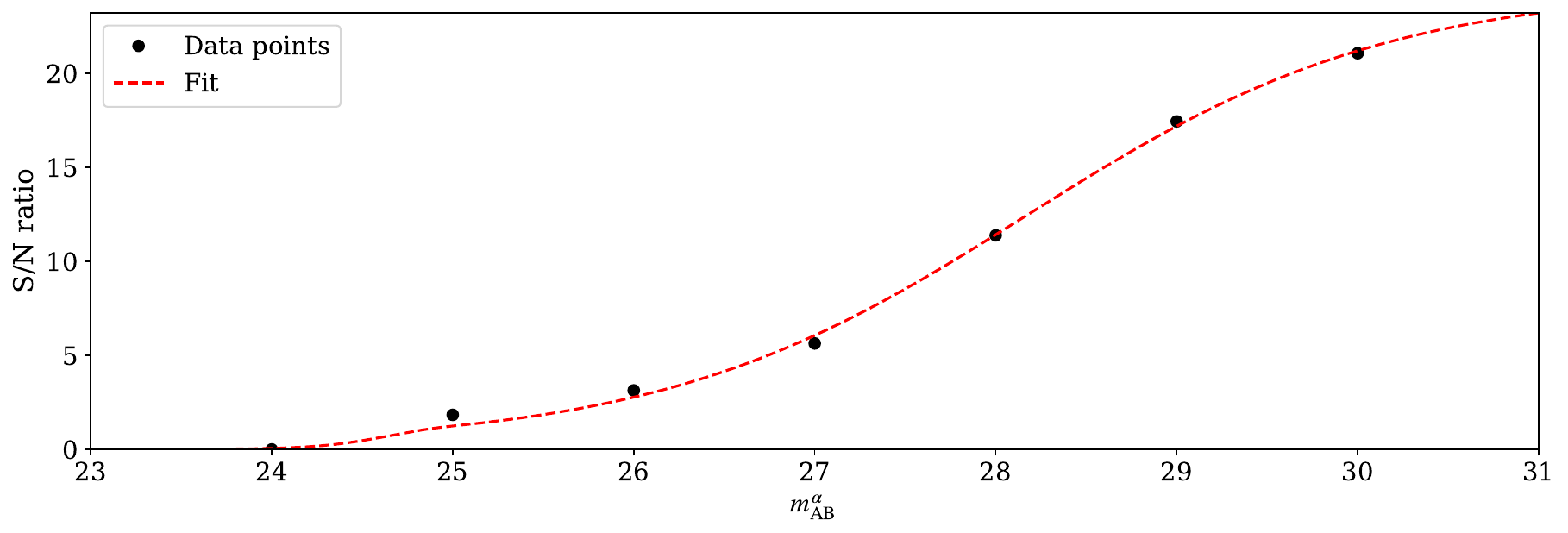}}
        \caption{The fit function for the signal-to-noise ratio as a function of observing depth for a spectroscopic survey with a field of view of $1$ square degree.}
        \label{fig:double_sigmoid}
    \end{figure}

    \noindent We chose this form since it obeys the limit of approaching zero as the limiting magnitude approaches zero (the limit where no galaxies are detected) as well as the limit wherein it approaches some maximum S/N as the limiting magnitude tends to infinity (the limit where all galaxies are detected). We chose to use a double sigmoid rather than a typical sigmoid since it tends to fit the data better. Figure \ref{fig:double_sigmoid} illustrates $\frac{\textrm{S}}{\textrm{N}}(m_{\textrm{AB}}|\delta_z)$for a spectroscopic Ly-$\alpha$ survey assuming normal 21-cm foregrounds and a field of view of $1$ square degree.

    \section{Increasing Field of View versus Observing Depth}
    \label{apdx:fov}

    The S/N in the cross-power spectrum increases when either the field of view of the galaxy survey or its limiting magnitude increases, as seen in Figures \ref{fig:depth-scaling}-\ref{fig:depth-scaling-fresco}. In designing a galaxy survey for the purpose of cross-correlation with the 21-cm signal, one asks: what is the better use of observing time? Increasing the total number of telescope pointings or increasing the integration time per pointing?

    Let us proceed with a very basic mathematical argument for favoring the field of view over the observing depth. Consider a survey of variable total observing time, $t$. The field of view of the survey scales linearly with $t$,

    \begin{equation}
        \textrm{FoV}\propto t.
    \end{equation}

    \noindent Meanwhile the observing depth (in magnitudes) scales logarithmically with $t$ via

    \begin{equation}
        m_{\textrm{AB}}\propto \ln(t).
    \end{equation}

    From Equation \ref{eq:modes} we know that for a fixed observing depth the S/N in the detection of the cross-power spectrum scales as the square root of the field of view, i.e.,

    \begin{equation}
        \hat{s}_\textrm{fixed depth}\propto \textrm{FoV}^{1/2}\propto t^{1/2}.
    \end{equation}

    For direct comparison, let us produce an analogous scaling relation for S/N with a fixed field of view and varying observing depth. In Appendix \ref{apdx:fn} we produced a functional form for the dependence of S/N on observing depth by fitting a double sigmoid to the results of our forward model. Here let us consider only the upper sigmoid, which is a simple sigmoid limited to the domain $m_{\textrm{AB}}\geq25$. Here,

    \begin{equation}
        \hat{s}_{\textrm{fixed FoV}}=h[1+\exp\{-k(m_{\textrm{AB}}-r)\}]^{-1}+c.
    \end{equation}

    Letting $m_{\textrm{AB}}=C\ln(t)+D$ and substituting this into the above equation, we obtain

    \begin{equation}
        \hat{s}_{\textrm{fixed FoV}}=h[1+t^{-kC}e^{kr-kD}]^{-1}+c.
    \end{equation}

    To evaluate whether increasing observing depth for fixed 
    field field of view or vice versa is a more efficient 
    use of observing time, let us evaluate the ratio $\hat{s}'_{\textrm{fixed FoV}}/\hat{s}'_{\textrm{fixed depth}}$, where $'$ indicates a partial derivative with respect to time. For a survey of fixed observing depth, the S/N scales via

    \begin{equation}
        \hat{s}_{\textrm{fixed depth}}=\frac{\hat{s}_0}{t_0^{1/2}}t^{1/2},
    \end{equation}

    \noindent where $\hat{s}_0$ is the S/N of some reference survey using the same instrument and $t_0$ is the time required to produce that survey. The partial derivative 
    of this equation with respect to time is 

    \begin{equation}
        \hat{s}'_{\textrm{fixed depth}}=\frac{\hat{s}_0}{t_0^{1/2}t^{1/2}}.
    \end{equation}

    \noindent Now let us take the partial derivative of $\hat{s}_{\textrm{fixed FoV}}$ to produce

    \begin{equation}
        \hat{s}'_{\textrm{fixed FoV}}=hkCe^{kr-kD}t^{-kC}[1+t^{-kC}e^{kr-kD}]^{-2}.
    \end{equation}

    \noindent Therefore, the ratio of these quantities is 

    \begin{equation}
        \mathcal{R}\equiv\frac{\hat{s}'_{\textrm{fixed FoV}}}{\hat{s}'_{\textrm{fixed depth}}}=\frac{t_0^{1/2}}{\hat{s}_0}hkCe^{kr-kD}t^{-kC-1/2}[1+t^{-kC}e^{kr-kD}]^{-2}.
    \end{equation}

    \noindent We now simplify the above expression by defining $\alpha=\frac{t_0^{1/2}}{\hat{s}_0}h$,
    $\beta=e^{kr-kD}$, and $\gamma=kC$, allowing us to write

    \begin{equation}
        \mathcal{R}=\alpha\beta\gamma t^{-\gamma-1/2}[1+\beta t^{-\gamma}]^{-2}.
    \end{equation}

    \noindent If $\gamma$ is positive, then $\mathcal{R}$ is positive and defined only for $t>0$ and has one critical point, a maximum. Recall that if $\mathcal{R}>1$, then increasing observing depth for a fixed field of view is more efficient than increasing field of view for a fixed observing depth; if $\mathcal{R}<1$, then the reverse is true. A simple way to see whether the former is ever true is to check whether the maximum value of $\mathcal{R}$ is greater than $1$.

    When we vary either the field of view or the observing depth, we vary them with respect to some fixed configuration of field of view, depth, $t_0$, and $\hat{s}_0$. This necessarily requires some assumptions about the survey configuration, including the instrument's limiting magnitude, field of view per pointing, galaxy selection criterion, 21-cm instrument, and 21-cm foregrounds. In what follows, let us consider perturbations of a spectroscopic survey which uses Lyman-$\alpha$-selected galaxies and is paired with a 21-cm measurement from the SKA (i.e., the experiment shown in the rightmost panel of Figure \ref{fig:depth-scaling}). The fixed point in our analysis shall consist of a survey covering $1$ square degree at a depth of $m_{\textrm{AB}}=26$. Such a survey yields a S/N of $\hat{s}_0=2.78$. The amount of time required to complete a survey depends on its telescope's field of view per pointing, $A_I$, and $5$-hour $5\sigma$ limiting magnitude, $m_{\textrm{AB},0}$ via the relation 

    \begin{equation}
        t = \frac{1}{A_I}10^{\frac{4}{5}(m_{\textrm{AB}}-m_{\textrm{AB},0})}, 
        \label{eq:time}
    \end{equation}

    \noindent which for $m_{\textrm{AB}}=26$ yields $t_0=A_I^{-1}10^{4(26-m_{\textrm{AB},0})/5}$.

    Having determined our reference survey, let us now find the critical point of $\mathcal{R}$. The partial derivative of $\mathcal{R}$ with respect to $t$ is

    \begin{equation}
        \mathcal{R}'=-\alpha\beta\gamma t^{-\gamma-3/2}[1+\beta t^{-\gamma}]^{-2}\times\left\{\left(\gamma+\frac{1}{2}\right)-2\beta\gamma t^{-\gamma}[1+\beta t^{-\gamma}]^{-1}\right\}.
    \end{equation}

    \noindent The maximum occurs when $\mathcal{R}'=0$, and solving the above yields a critical time of

    \begin{equation}
        t_c=\left[\frac{1}{\beta}\left(\frac{2\beta\gamma}{\gamma+\frac{1}{2}}-1\right)\right]^{-1/\gamma}.
    \end{equation}

    \noindent Substitution into the equation for $\mathcal{R}$ reveals a maximum value of 

    \begin{equation}
        \mathcal{R}_{\textrm{max}}=\frac{\alpha}{4\beta\gamma}\left(\gamma+\frac{1}{2}\right)^2\left[\frac{1}{\beta}\left(\frac{2\beta\gamma}{\gamma+\frac{1}{2}}-1\right)\right]^{1+2/\gamma}.
    \end{equation}

    Now let us evaluate the above for realistic values of $\alpha$, $\beta$, and $\gamma$.
    $\alpha$ depends on $t_0$, $\hat{s}_0$, and $h$. $t_0$ depends on the field of view and limiting  magnitude of the specific instrument used for the survey while $\hat{s}_0$ and $h$ are fixed by the redshift uncertainty, galaxy selection criterion, 21-cm instrument, and 21-cm foregrounds assumed in the model. For our SKA-spectroscopic survey example, $\hat{s}=2.78$ and $h=24.34$, giving us

    \begin{equation}
        \alpha=8.75A_I^{-1/2}10^{2(26-m_{\textrm{AB},0})/5}.
    \end{equation}

    \noindent Next we have $\beta$, which depends on $k$, $r$, and $D$. $k=0.99$ and $r=28.17$ are fixed by the same criteria as $h$ whereas $D$ depends on our choice of instrument. Rearranging Equation \ref{eq:time} yields

    \begin{equation}
        m_{\textrm{AB}}=\frac{5}{4\ln10}\ln t + \frac{5}{4\ln10}\ln A_I + m_{\textrm{AB},0},
    \end{equation}

    \noindent wherein $D=\frac{5}{4\ln10}\ln A_I + m_{\textrm{AB},0}$ and $C=\frac{5}{4\ln10}$ with reference to our aforementioned form for $m_{\textrm{AB}}(t)$. Therefore,

    \begin{equation}
        \beta = A_I^{-0.54}\exp\{0.99(29.17-m_{\textrm{AB},0})\}.
    \end{equation}

    \noindent Finally,

    \begin{equation}
        \gamma=0.54.
    \end{equation}

    Substituting these relations into the equation for $\mathcal{R}_{\textrm{max}}$ yields

    \begin{multline}
        \mathcal{R}_{\textrm{max}}=2.32 A_I^{0.03}\cdot10^{2(26-m_{\textrm{AB},0})/5}e^{0.99(m_{\textrm{AB},0}-28.17)}\\
        \times[1.03A_I^{0.03}-A_I^{0.53}e^{0.99(m_{\textrm{AB},0}-28.17)}]^{4.77}.        
    \end{multline}

    Now we have an equation which depends only on the field of view per pointing 
    and limiting magnitude of an instrument. Varying these allows us to explore 
    various possible spectroscopic survey instruments, as long as they select 
    for Lyman-$\alpha$ emission and are cross-correlated with a 21-cm detection
    made using SKA-low.

        \begin{figure}
            \centering
            {\includegraphics[width=\textwidth]{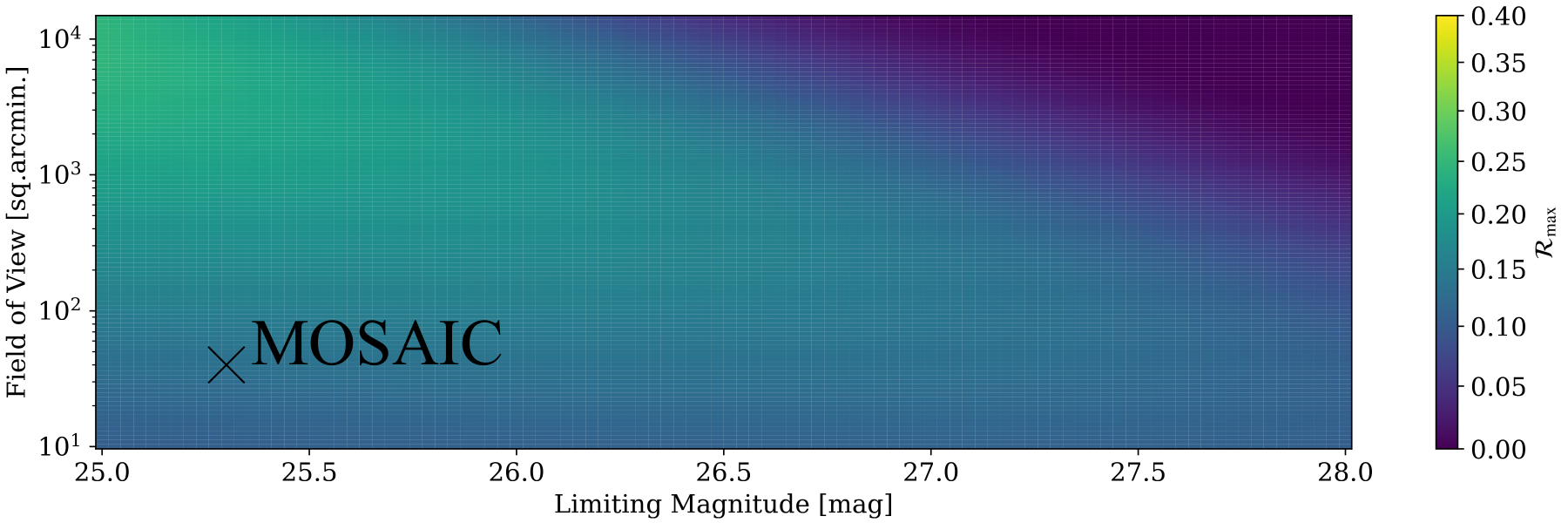}}
            \caption{The maximum value of the ratio between the 
            signal gain per unit observing time for fixed field of view to 
            the same for fixed observing depth, calculated for various 
            telescope fields of view and limiting magnitudes. 
            All values are calculated with reference to a Lyman-$\alpha$-selected galaxy survey with spectroscopic redshifts used for cross-correlation with 
            a 21-cm measurement from SKA-low, assuming that 21-cm foregrounds 
            extend to the horizon. Values greater than one would indicate that 
            increasing the observing depth yields more signal than increasing the 
            field of view, for the same amount of time spent. All values in the 
            domain shown are less than one, indicating that increasing the 
            field of view yields more signal per unit time than increasing 
            the observing depth.}
            \label{fig:rmax}
        \end{figure}

    Figure \ref{fig:rmax} summarizes the results of our investigation. We find that for 
    all values of $A_I\in(25,1000)$ square arcminutes and $m_{\textrm{AB},0}\in(25,28)$, 
    it is more time-efficient to increase the field of view of the survey rather than
    increasing its depth, relative to a fiducial survey which covers $1$ square degree
    with a depth of $m_{\textrm{AB}}=26$. Therefore, we say that for any realistic 
    spectroscopic survey, the optimal survey design for the purposes of maximizing 
    the signal in the cross-power spectrum is to target a depth of $m_{\textrm{AB}}\sim26$
    and cover as much contiguous area on the sky as possible.

\end{appendix}

\end{document}